\documentstyle[12pt]{article}
\input epsf
\begin{document}
\bibliographystyle{unsrt}
\def\question#1{{{\marginpar{\small \sc #1}}}}

\rightline{hep-ph/9610280}
\rightline{RAL-96-052}
\rightline{RU-96-35}  
\baselineskip=18pt
\begin{center}
{\bf \LARGE Determining the Gluonic Content of Isoscalar Mesons}\\
\vspace*{0.9in}
{\large Frank E. Close}\footnote{\tt{e-mail: fec@v2.rl.ac.uk}} \\ 
\vspace{.1in}
{\it Rutherford Appleton Laboratory}\\
{\it Chilton, Didcot, OX11 0QX, England}\\ 
\vspace{0.1in} 
{\large Glennys R. Farrar}\footnote{\tt{e-mail: farrar@farrar.rutgers.edu}} \\ 
{\it Department of Physics and Astronomy}\\
{\it Rutgers University, Piscataway, NJ 08855, USA}\\
\vspace*{0.1in} 
{\large Zhenping Li}\footnote{\tt{e-mail: zpli@ibm320h.phy.pku.edu.cn}} \\
{\it Physics Department}\\ 
{\it Peking University, Beijing, 100871, P. R. China}\\
\end{center}

\begin{abstract}
We develop tools to
determine the gluonic content of a resonance of 
known mass, width and $J^{PC}$ from its branching fraction in radiative
quarkonium decays and production cross section in $\gamma \gamma$ 
collisions.  We test the procedures by applying them to known $q\bar{q}$
mesons, then analyze four leading glueball candidates.  We identify
inconsistencies in data for $J/\psi \to \gamma f_0(1500)$ 
and $J/\psi \to \gamma f_J(1710)$ whose resolution can quantify their
glueball status.When $\Gamma(f_0(1500) \to \gamma \gamma )$ and
$\Gamma(f_J(1710) \to \gamma \gamma)$ are known, the $n\bar{n},
s\bar{s},gg$ mixing angles can be determined.  The enigmatic situation
in 1400-1500 MeV region of the isosinglet $0^{-+}$ sector is discussed. 
\end{abstract}

\newpage

\section {Introduction}

\hspace*{2em}There has been considerable recent interest in the
possible sighting of glueballs. Four states are of particular
interest:   
\begin{itemize}
\item 
$f_0(1500)$\cite{lear,cafe95,bugg}
\item
$f_J(1710)$\cite{weing} where $J=0$ or $2$\cite{pdg94}
\item $\xi(2230)$\cite{beijing} 
\item $\eta(1440)$\cite{ishikawa}, now resolved into two pseudoscalars.
\end{itemize}
In this paper we calculate the production rate of conventional mesons
($q\bar{q}$) and glueballs (``$G$") in the radiative decay of vector
quarkonium, as a function of their mass, angular momentum, and width.
If the data on the radiative production of these states are correct, we
find that     

(i) The $f_0(1500)$ is probably produced at a rate too high to be a
$q\bar{q}$ state.  The average of world data suggests it is a
glueball-$q \bar{q}$ mixture.  

(ii) The $f_J(1710)$ is produced at a rate which is consistent with
it being $q\bar{q}$, only if $J=2$.  If $J=0$, its production
rate is too high for it to be a pure $q\bar{q}$ state but is consistent
with it being a glueball or mixed $q \bar{q}$-glueball having a large
glueball component. 

(iii) The $\xi(2230)$, whose width is $\sim 20$ MeV, is produced at a
rate too high to be a $q\bar{q}$ state for either $J=0$ or $2$.  If
$J=2$, it is consistent with being a glueball.  The assignment $J=0$
would require $Br(J/\psi \rightarrow \gamma \xi) \,\raisebox{-0.13cm}{$\stackrel{\textstyle<}{\textstyle\sim}$}\, 3 ~10^{-4}$,
which already may be excluded. 

(iv) The enhancement once called $\eta(1440)$ has been resolved
into two states.  The higher mass $\eta(1480)$ is dominantly
$s\bar{s}$ with some glue admixture, while the lower state
$\eta(1410)$ has strong affinity for glue.

\noindent We note what improvements in data would allow these
constraints to be sharpened.  We also analyze 3-state mixing in the
$0^{++}$ sector between $n \bar{n}$, $s \bar{s}$ and $gg$, showing how
$\Gamma(J/\psi \rightarrow \gamma f_0)$ and $\Gamma(f_0 \rightarrow
\gamma \gamma)$ determine mixing angles. 

The interest in these states as glueball candidates is motivated on
both phenomenological and theoretical grounds.  Phenomenologically,
these states satisfy qualitative criteria expected for
glueballs\cite{closerev}:
\begin{enumerate}
\item
Glueballs should be favoured over ordinary mesons in the
central region of high energy scattering processes, away from beam and
target quarks.  The $f_J(1710)$ and possibly the $f_0(1500)$ have been
seen in the central region in $pp$
collisions\cite{Kirk,Gentral}.
\item
Glueballs should be produced in
proton-antiproton annihilation, where the destruction of quarks 
creates opportunity for gluons to be manifested.  This is the Crystal
Barrel \cite{Anis}, and E760 \cite{Hasan1}
production mechanism,
 in which detailed decay systematics of
$f_0(1500)$ have been studied. The empirical situation with regard to
$f_J(1710)$ and $\xi(2230)$ is currently under investigation. The
$\eta(1440)$ is clearly seen in $p\bar{p}$ annihilation\cite{obelixE,cbiota}
\item
Glueballs should be enhanced compared to ordinary mesons in radiative
quarkonium decay.  In fact, all four of these resonances are produced
in radiative $J/\psi$ decay at a level typically of $\sim1$ part per
thousand. A major purpose of this paper is to decide whether these rates
indicate that these resonances are glueballs, or not. 
\end{enumerate}

On the theoretical side, lattice QCD predicts that the lightest
``ideal" (i.e., quenched approximation) glueball be $0^{++}$, with
state-of-the-art mass predictions of $1.55 \pm 0.05$ GeV\cite{ukqcd}
and $1.74 \pm 0.07$ GeV\cite{weing}.  That lattice QCD is now
concerned with such fine details represents considerable advance in
the field and raises both opportunity and enigmas. First, it
encourages serious consideration of the further lattice predictions
that the $2^{++}$ glueball lie in the $2.2$ GeV region, and hence
raises interest in the $\xi(2230)$.  Secondly, it suggests that scalar
mesons in the $1.5-1.7$ GeV region merit special attention.  Amsler
and Close\cite{cafe95} have pointed out that the $f_0(1500)$ shares
features expected for a glueball that is mixed with the nearby
isoscalar members of the $^3P_0$ $q\bar{q}$ nonet.  If the $f_J(1710)$
proves to have $J=2$, then it is not a candidate for the ground state
glueball and the $f_0(1500)$ will be essentially unchallenged.
On the other hand, if the $f_J(1710)$ has $J=0$ it becomes a potentially
interesting glueball candidate.  Indeed, Sexton, Vaccarino and
Weingarten\cite{weinprl} argue that $f_{J=0}(1710)$ should be
identified with the ground state glueball, based on its similarity in
mass and decay properties to the state seen in their lattice
simulation.  While the consistency between theoretical mass
predictions and the observed states is quite satisfactory in the
$0^{++}$ and $2^{++}$ sectors, this is not the case in the $0^{-+}$
sector.  Both lattice and sum rule calculations place the
lightest $0^{-+}$ glueball at or above the $2^{++}$ glueball so
that the appearance of a glueball-like pseudoscalar in the 1.4-1.5 GeV
region is unexpected.  It is interesting that its properties are
consistent with those predicted for the gluino-gluino bound state in
supersymmetry breaking scenarios with a light gluino\cite{gluino}

In order to make quantitative estimates of the gluonic content of
isosinglet mesons, we use their measured radiative quarkonium production
rates and gamma-gamma decay widths.  We apply the relationship
proposed by Cakir and Farrar\cite{cak} (CF) between the branching
fraction for a resonance $R$ in radiative quarkonium decay, $b_{rad}({Q \bar{Q}}_V \rightarrow \gamma +R) \equiv
\Gamma({Q \bar{Q}}_V \rightarrow \gamma + X)$ and its branching
fraction to gluons, $br(R \rightarrow gg) \equiv \Gamma (R \rightarrow gg ) /
\Gamma(R\rightarrow \rm{all})$:  
\begin{equation} 
\label{CF}
b_{rad}(Q\bar Q_V\to \gamma +R_J)=
\frac {c_Rx|H_{J}(x)|^2}{8\pi(\pi^2-9)}\frac{m_R}{M_V^2}\Gamma_{tot} br(R_J
\rightarrow gg),
\end{equation}
where $M_V$ and $m_R$ are masses of the initial and final resonances, and
$x \equiv 1-\frac {m_R^2}{M^2_V}$; $c_R$ is a numerical factor and
$H_J(x)$ a loop integral which will be discussed in section 2.  For a
resonance of known mass, total width ($\Gamma_{tot}$), and $J^{PC}$, a
relationship such as eq. (\ref{CF}) would determine $br(R \rightarrow gg)$ if 
$b_{rad}({Q \bar{Q}}_V \rightarrow \gamma +R)$ were known.  CF argued that one expects    
\begin{equation}
\begin{array}{lcl}
br(R[q \bar{q}] \rightarrow gg)& =& 0(\alpha^2_s) \simeq 0.1-0.2\nonumber\\
br(R[G] \rightarrow gg)& \simeq& O(1).\nonumber\\
\end{array}
\end{equation}
Thus knowledge of $br(R \rightarrow gg)$ would give quantitative information on the
glueball content of a particular resonance.  Using $H_J(x)$ determined in
the non-relativistic quark model (NRQM), CF found that known $q\bar{q}$
resonances (such as $f_2$(1270)) satisfy the former and noted that the
$f_0(1710)$ might be an example of the latter.  

In the present paper we give a more general discussion of the
functions $H_J(x)$ needed to employ Eq. (\ref{CF}),
clarify some of the assumptions implicit in its derivation, and
verify that application of the relation does not depend on
flavor mixing.  A number of experimental tests are proposed.  We
discuss the additional information that can be obtained when the
cross section for production of the resonance in $\gamma \gamma$
collisions is  
known. 

The paper is organized as follows.  We start with a section on
the formalism and its model dependence (sec. \ref{theory}). 
A general treatment of the problem
requires defining form factors for the coupling of a resonance, of
specified $J^{PC}$, to a pair of virtual gluons.  The partial width
$\Gamma(R \rightarrow gg)$ fixes a linear combination of the form factors at the
on-shell-gluon point.  The internal structure of the resonance
determines both the relative size of the various form factors at the
on-shell-gluon point, as well as their virtuality dependence, just as
in the case of the nucleon electromagnetic form factors $F_1$ and
$F_2$.  The $H_J(x)$'s depend on integrals of the form factors over the
gluon virtualities.  In sec. \ref{hoc} we discuss higher order
corrections, scale dependence, and the relationship of the $R \rightarrow
g g$ form factors to the $R \rightarrow \gamma \gamma$ amplitudes
which can in principle be measured in a photon-photon collider.
(The phenomenology of the latter is developed in section \ref{gam2}).
In sec. \ref{result} we re-express eq. (\ref{CF})
so that its implications are more transparent and it is easier to
apply to data.  Our central results (eqs. \ref{0++}-\ref{0-+}) show
how the spin of a resonance and its width into gluons fixes its
production rate in radiative quarkonium decay. In section 3.2 we show
that the relations do not depend on flavor mixing and that the known
$q\bar{q}$ resonances $f_2(1270;~1525)$ satisfy eq.(2). In sec.
3.3 we discuss the utility of experimental study of radiative upsilon
decay, especially $\Upsilon \to \gamma \chi$; in sec. 3.4 we
investigate $1^{++}$ mesons. In secs. \ref{ss:f1500}-\ref{ss:xi2330}
we apply eqs. (\ref{0++}-\ref{0-+}) to the $f_0(1500),~f_J(1710)$,
$\xi(2300)$, and $\eta(1410;1480)$ leading to the results listed at
the beginning of the introduction.  In section 5 we discuss how
$\gamma \gamma \to R$ in combination with $J/\psi \to \gamma R$ can
help to distinguish glueballs from $q\bar{q}$ states and determine
basic parameters. Section 6 considers the possibility of
glueball-$q\bar{q}$ mixing involving three states, $f_0(1370),
f_0(1500)$ and $f_{0?}(1710)$.  In general we use the
nomenclature of the 1994 edition of the Particle Data Tables
throughout\cite{pdg94}. Readers mainly interested in the
phenomenological results can proceed directly to sections \ref{result}
et seq. 

\section{Formalism}
\label{theory}

\subsection {$b_{rad}({Q \bar{Q}}_V \rightarrow \gamma +R)$}
\hspace*{2em}
The decay width for the radiative decay of a vector heavy quarkonium
state, $ Q\bar Q_V\to \gamma R$ is 
\begin{equation}\label{QQgamRgen}
\Gamma = \frac 1{24\pi} \frac {k}{M_V^2} \sum_{i,f}|A|^2,
\end{equation}
where $k$ is the photon momentum, $M_V$ is the  mass of the spin-1
$Q\bar Q_V$ state, and the summation is over the polarizations of the
initial and final particles.  If the resonance $R$ does not contain a
``valence'' $Q \bar{Q}$ component, the decay occurs through a two
gluon intermediate state, in leading order pQCD, and the amplitude $A$
is given by
\begin{equation}\label{QQgamR}
A= \frac 12\sum \int \frac {d^4k}
{(2\pi)^4} \frac 1{k_1^2} \frac 1{k_2^2} <(Q\bar Q)_V | \gamma g^ag^b> 
<g^ag^b | R>.
\end{equation}
The summation is over the polarization vectors $\epsilon_{1,2}$ and
color indices $a,b$ of the intermediate gluons, whose momenta are
denoted $k_{1,2}$.  The amplitude $<Q\bar Q_V | \gamma gg>$ 
couples a vector $Q\bar Q$ state with polarization and momentum
($E,K$) to a photon ($\epsilon, k$) and the two virtual gluons.  For
heavy quarks, $Q$, this amplitude is reliably given by perturbative
QCD.  Using the non-relativistic quark model to describe the $Q\bar
Q_V$ wavefunction (the pQCD-NRQM
approximation\cite{korner,bill,koller}):    
\begin{equation}\label{QQgamgg}
<Q\bar Q_V | \gamma g^a g^b>= e_Q g_s^2 \delta^{ab}\sqrt{\frac 23} 
\frac{iR_V(0)}{\sqrt{4\pi M^3_V}} 
\frac {M^2_V}{k\cdot(k_1+k_2)k_1\cdot(k+k_2)k_2\cdot(k+k_1)}a_V,
\end{equation}
where 
\begin{eqnarray}\label{aV}
a_V= \epsilon_1\cdot \epsilon_2[-k_1\cdot k\epsilon\cdot k_2E\cdot 
k_1-k_2\cdot k\epsilon\cdot k_1 E\cdot k_2-k_1\cdot k k_2\cdot k 
E\cdot\epsilon]   \nonumber \\
+E\cdot \epsilon [k_1\cdot k\epsilon_1\cdot k_2\epsilon_2\cdot k+k_2\cdot 
k\epsilon_2\cdot k_1\epsilon_1\cdot k-k_1\cdot k_2\epsilon_1\cdot 
k\epsilon_2\cdot k]\nonumber \\
+\{ \epsilon_1,k_1 \Leftrightarrow \epsilon,k\} + \{ \epsilon_2, k_2 
\Leftrightarrow \epsilon,k\}.
\end{eqnarray}
$R_V(0)$ is the $(Q\bar{Q})_V$ wavefunction at the origin and $e_Q$ is
the charge of the heavy quark $Q$.

The amplitude $<g^a(k_1,\epsilon_1)g^b(k_2,\epsilon_2) | R>$ must be
linear in $\epsilon_1$ and $\epsilon_2$ and Lorentz and gauge
invariant.  A linearly independent set of tensor structures satisfying
these requirements for $J^{++}$ states is given in \cite{kuhn}.
Thus we can write, with the shorthand $G_{\mu\nu} \leftrightarrow
k_{\mu} \epsilon_{\nu} - \epsilon_{\mu} k_{\nu}$ and the convention
that $G^{1(2)}$ refers to $k_{1(2)},\epsilon_{1(2)}$, and suppressing
$J^{PC}$ labels on the $F_i$'s: 
\begin{eqnarray}
<g^a g^b | 0^{++}> & = &  \delta^{ab} \frac
{A_{0^{++}}P_{\rho\sigma}}{\sqrt{3}}  \left [
F_1(k_1^2,k_2^2)G_{\mu\rho}^1 G_{\nu\sigma}^2  
+ F_2(k_1^2,k_2^2) k_1^{\mu}G^1_{\mu\rho}G^2_{\nu\sigma}k_2^{\nu}
\right ],
\label{ff0++} \\
<g^a g^b | 0^{-+}> & = & \delta^{ab} A_{0^{-+}}  F_1(k_1^2,k_2^2) 
\epsilon^{\mu\nu\rho\sigma}
G_{\mu\nu}^1G_{\rho\sigma}^2,
\label{ff0-+} \\  
<g^a g^b | 1^{++}> & = & \delta^{ab} A_{1^{++}} \left( F_1(k_1^2,k_2^2)
\epsilon^{\mu\nu\rho\sigma} \epsilon_\sigma (G_{\mu\nu}^1 G_{\rho\lambda}^2
k_2^\lambda  +  G_{\mu\nu}^2 G_{\rho\lambda}^1 k_1^\lambda ) \right. 
\nonumber \\
& & \left. + F_2(k_1^2,k_2^2) \epsilon^{\mu\nu\rho\sigma}\epsilon_\alpha
(k_1^{\alpha}-k_2^\alpha)G^1_{\mu\nu}G^2_{\rho\sigma}\right),
\label{ff1++} \\
<g^a g^b | 2^{++}> & = & \delta^{ab} A_{2^{++}} \epsilon_{\rho\sigma} 
 \left [ F_1(k_1^2,k_2^2)G^1_{\mu\rho}G^2_{\mu\sigma}  
 + F_2(k_1^2,k_2^2) k_1^{\rho}k_2^{\sigma}
 G_{\mu\nu}^1G_{\mu\nu}^2 \right. + 
\nonumber \\
& &  F_3(k_1^2,k_2^2) k_1^{\mu}G^{1}_{\mu\rho} G^{2}_{\nu\sigma}k_2^{\nu} 
\left. + F_4(k_1^2,k_2^2) k_1^{\rho}k_2^{\sigma}
k_1^{\mu}G^1_{\mu\rho}G^2_{\nu\rho}k_2^{\nu}\right ], 
\label{ff2++}
\end{eqnarray}
where 
\begin{equation}\label{pmunu}
P_{\rho\sigma} \equiv g_{\rho\sigma}-\frac {P_{\rho}P_{\sigma}}{m^2},
\end{equation}
for a resonance with mass $m$ and momentum $P_{\mu}$.  
Here $\epsilon_\rho$ and $\epsilon_{\rho\sigma}$
are the polarization vector and tensor for a vector or tensor
resonance, and satisfy the relations
\begin{eqnarray}\label{vector}
\sum_{\epsilon} \epsilon_\rho\epsilon_\sigma & = & g_{\rho\sigma}-
\frac {P_{\rho}P_{\sigma}}{m^2}\quad {\rm or} \nonumber \\
\sum_{\epsilon}
\epsilon_{\rho\sigma}\epsilon_{\rho^\prime\sigma^\prime} &
= & \frac
12(P_{\rho\rho^\prime}P_{\sigma\sigma^\prime}+P_{\rho\sigma^\prime}
P_{\sigma\rho^\prime})-\frac
13P_{\rho\sigma}P_{\rho^\prime\sigma^\prime}.
\end{eqnarray}

The resonance $R$ could be $q\bar{q}$, glueball or mixture of both.
The form factors $F_i(k_1^2,k_2^2)$ depend on the composition of $R$.
This will be discussed below.  We adopt the normalization convention
that $F_1[J^{PC}](0,0) = 1$.  Also, we use the shorthand for
on-shell form factors $F_i[J^{PC}](0,0) = F_i[J^{PC}]$.  Note that for
$k_1^2 = k_2^2 = 0$, $<g^a g^a | 1^{++}> = 0$ as it must by Furry's
theorem.  The constants $A_{J^{PC}}$ in Eqs. \ref{ff0++}-\ref{ff2++}
are dependent on the coupling between gluons and the resonance
constituents, the wavefunction of the resonance and its mass
(specific examples for $q\bar{q}$ are given below in eq. 
(24)).

After summing over the color index of the final two gluon state, the 
general expression for the total  width of a resonance R 
decaying into two real gluons\cite{pdg94} is 
\begin{equation}\label{gamma}
\Gamma (R\to gg)=\frac 1{(2J+1)}\frac
1{2m\pi}\sum_{\epsilon_1,\epsilon_2} |<gg|R>|^2, 
\end{equation}
where $<gg|R>$ is the coefficient of $\delta^{ab}$ in eqs.
\ref{ff0++}-\ref{ff2++}.  For example in the case $R=0^{++}$, using
the matrix element $<gg|0^{++}>$ given by eq. \ref{ff0++}, and summing
over gluon polarizations $\epsilon_{1,2}$, gives
\begin{eqnarray}\label{gamma1}
\sum_{\epsilon_{1,2}}|<gg|0^{++}>|^2 & = & \frac 13|A_{O^{++}}|^2\left [
(k_1\cdot k_2)^2 g_{\mu\mu^\prime}g_{\nu\nu^\prime}
+k_1^{\mu}k_1^{\nu}k_2^{\mu^\prime}k_2^{\nu^\prime} \right . \nonumber \\
& & +k_1^{\mu^\prime}k_1^{\nu^\prime}k_2^{\mu}k_2^{\nu}
-2k_1\cdot k_2(g_{\mu\mu^\prime}k_1^{\nu}k_2^{\nu^\prime}
+g_{\nu\nu^\prime}k_1^{\mu^\prime}k_2^{\mu}) \nonumber \\ 
& & + \left .   k_1\cdot k_2 (g_{\mu\nu}k_1^{\mu^\prime}
k_2^{\nu^\prime}+g_{\mu^\prime\nu^\prime}k_1^{\mu}
k_2^{\nu})
\right ]P_{\mu\nu}
P_{\mu^\prime\nu^\prime} \nonumber \\
& = & \frac 38 m^4|A_{O^{++}}|^2,
\end{eqnarray}
which leads to 
\begin{equation}\label{gam0++}
\Gamma(R_{0^{++}}) = \frac {3m^3}{16\pi}|A_{0^{++}}|^2.
\end{equation}

The decay widths for the other states are obtained by the same
procedure and read

\begin{equation}\label{gam0-+}
\Gamma(R_{0^{-+}}) = \frac {2m^3}{\pi}|A_{0^{-+}}|^2, 
\end{equation}
and
\begin{equation}\label{gam2++}
\Gamma(R_{2^{++}}) = \frac {m^3}{20\pi}|A_{2^{++}}|^2\left (1+
\frac {m^4}{12} F_2^2[2^{++}]\right ).
\end{equation}
While these are nominally two-gluon widths, when the scale of
resolution of the gluons is taken large enough (see ref.\cite{cak} and
below) they become the total gluonic widths.  Note also that since
$A_{1^{++}}$ cannot be fixed in this way, we focus primarily on
$0^{++}$, $2^{++}$, and $0^{-+}$ states.  We return to the axial
mesons in sec. \ref{axials}

Given the form factors appearing in $<g^ag^b | R>$ and the NRQM-pQCD
formula (eq. \ref{QQgamgg}) for $<Q\bar Q_V |\gamma g^ag^b>$, the
integral $\int d^4k \frac 1{k_1^2k_2^2} <Q\bar Q_V |\gamma
g^ag^b><g^ag^b | R>$ can be carried out, thus determining the $H_J(x)$'s
appearing in eq. (\ref{CF}).  The analysis of
ref.\cite{cak} assumed that the relative size of the on-shell form
factors and their dependence on gluon virtualities is universal, for
heavy and light $q \bar{q}$  mesons and for glueballs. There is no
general reason why this should be the case. 
For example, we know from form factors of electromagnetic
and weak currents that some aspects of form factors are universal
\footnote{For instance, the leading $Q^2$ dependence of the nucleon and meson 
electromagnetic form factors depends only on the number of valence
constituents\cite{bf}.} while other aspects such as  the relative
magnitude of the nucleon on-shell form factors depend on detailed
structure of the bound state, in particular the constituent quark
magnetic moments.   
The next sections
describe the information we presently have on the $<gg|R>$ form factors.

\subsection{$<g g| R>$ form factors}
\label{ff}
\hspace*{2em}
A particular example for the $<gg|R>$ form factors is the case of 
$R=q\bar q$, where the quantity $<gg|R>$ has been 
modeled\cite{korner,koller,kuhn} as a QCD 
analogue\cite{barb} of the two photon coupling to 
positronium\cite{ale}.  In the NRQM approximation\cite{kuhn}
\begin{equation}\label{qqbar0++}
<gg|0^{++}(q\bar q)>=c^\prime \sqrt{\frac 16}
 \left [ G_{\mu\nu}^a G_{\mu\nu}^a(m^2+k_1\cdot
k_2)-2 k_1^{\nu}G_{\mu\nu}^a G_{\mu\rho}^a k_2^{\rho}\right ]
/(k_1\cdot k_2)^2,
\end{equation}
\begin{equation}
<gg | 1^{++}(q\bar q)>=c^\prime m\frac 12
 \epsilon^{\mu\nu\rho\sigma} \epsilon_\sigma \left [G_{\mu\nu}^1 
G_{\rho\lambda}^2
k_2^\lambda  +  G_{\mu\nu}^2 G_{\rho\lambda}^1 k_1^\lambda 
 \right ]
/(k_1\cdot k_2)^2,
\end{equation}
and 
\begin{equation}\label{17}
<gg | 2^{++}(q\bar q)>=c^\prime \sqrt{2}m^2 G_{\mu\rho}^a G_{\nu\rho}^a
e^{\mu\nu}/(k_1\cdot k_2)^2,
\end{equation}
where the constants $c^\prime$ are proportional to the derivative of
the radial wavefunctions at the origin:
\begin{equation}\label{18}
c^\prime=g_s^2\sqrt{\frac 1{m^3\pi}} R^\prime(0).
\end{equation}
Analogously the two gluon coupling for a $0^{-+}$ state is\cite{kuhn}
\begin{equation}\label{qqbar0-+}
<gg|0^{-+}(q\bar q)>=c \epsilon_{\rho\sigma\mu\nu}G^a_{\rho\sigma}
G^a_{\mu\nu}/k_1\cdot k_2,
\end{equation}
where 
\begin{equation}\label{18b}
c = g_s^2\frac 14\sqrt{\frac 1{3m\pi}}R(0).
\end{equation}
The above are particular models for Eqs. \ref{ff0++}-\ref{ff2++}.
To see the correspondence between the quantities $<gg|J^{PC} (q\bar
q)>$ and Eqs. \ref{ff0++}-\ref{ff2++}, it is convenient to note that  
Eq. \ref{qqbar0++} can be rewritten in the form of Eq. \ref{ff0++}:
\begin{equation}\label{qq0++}
<gg|0^{++}(q\bar q)>=c^\prime m^2\sqrt{\frac 23}
  G_{\alpha\mu}^{1a} G_{\alpha\nu}^{2a}P_{\mu\nu}  
/(k_1\cdot k_2)^2,
\end{equation}

The constants $A_{J^{PC}}$ in our Eqs. \ref{ff0++}-\ref{ff2++} are thus related
to the constants $c$ and $c^\prime$ of ref.\cite{kuhn} (Eqs. \ref{18} and
\ref{18b} above) by 
\begin{eqnarray}\label{A}
A_{0^{++}}=A_{2^{++}}=\sqrt{8}mA_{1^{++}}=\frac {4\sqrt{2}c^\prime}{m^2} 
\nonumber \\
A_{0^{-+}}=\frac {2c}{m^2},
\end{eqnarray}
and the matching  implies that the only non-zero form factors for
$<gg|q\bar q>$ in the NRQM are
\begin{equation}\label{2222}
F_1[0^{++}]=F_1[2^{++}]=F_1[1^{++}]=\frac {m^4}{4(k_1\cdot k_2)^2},
\end{equation}
and\cite{bg} 
\begin{equation}
F_1[0^{-+}]=\frac {m^2}{2k_1\cdot k_2}.
\end{equation}

Substituting Eq. \ref{A} into Eqs. \ref{gam0++}-\ref{gam2++}, gives
\begin{equation}
\Gamma((q\bar q)_{0^{++}})=96 \frac {\alpha_s^2}{m^4} |R^\prime (0)|^2,
\end{equation}
\begin{equation}\label{qq2++}
\Gamma((q\bar q)_{2^{++}})=\frac {128}5 \frac {\alpha_s^2}{m^4} 
|R^\prime (0)|^2,
\end{equation}
and
\begin{equation}
\Gamma((q\bar q)_{0^{-+}})=\frac 83 \frac {\alpha_s^2}{m^2} |R(0)|^2,
\end{equation}
which agree with those in Ref. \cite{cak}.

The gluon structure appearing in the form factors for the scalar and
tensor NRQM mesons (e.g., $G^a_{\alpha\mu}G^a_{\alpha\nu} P_{\mu \nu}$)
has been widely employed also for scalar and tensor
glueballs\cite{qsum}.
It can be considered a natural relativistic generalization of TE mode
glueballs in a cavity approximation such as the MIT bag
model\cite{barnes}:
\begin{equation}
\label{extra1}
\psi(G_{J^{++}}) = \langle 1 \alpha; 1 \beta | J, \alpha
+ \beta \rangle (\vec{\epsilon_1} \times \hat {\vec{k_1}})^{(\alpha)} 
(\vec{\epsilon_2} \times \hat{\vec{k_2}})^{(\beta)} \phi(r)
\end{equation}
where $\phi(r)$ is a radial wavefunction, and the superscripts
$\alpha~,\beta$ specify the projection of the angular momenta along
$\hat{z}$.   The relativistic generalization adopted in ref.
\cite{qsum} produces 
\begin{equation}\label{extra3}
\psi(G_{0^{++}})= \frac 1{\sqrt{3}} P_{\mu\nu}
 \frac {G^{1a}_{\mu\rho} G^{2a}_{\nu\rho}}{\sqrt {8} k_1\cdot k_2}  \phi(x)
\end{equation}
for the scalar and
\begin{equation}\label{extra4}
\psi(G_{2^{++}})= \epsilon_{\mu\nu}
 \frac {G^{1a}_{\mu\rho} G^{2a}_{\nu\rho}}{\sqrt {8} k_1\cdot k_2}  \phi(x)
\end{equation}
for the tensor states, where $P_{\mu\nu}$ and $\epsilon_{\mu\nu}$ are defined
in eqs. \ref{pmunu} and \ref{vector}.  Note that in cavity
approximation the same function $\phi(x)$ appears for both $0^{++}$
and $2^{++}$ states, so that the relative magnitudes of their form
factors are fixed.  The resulting matrix elements for the two gluon  
couplings of the glueballs are 
\begin{eqnarray}\label{extra5}
<gg|G(J^{PC})>  = Af(k_1^2,k_2^2){\cal P}^J_{\mu\nu}\frac
{G^{1a}_{\mu\rho} G^{2a}_{\nu\rho}}{k_1\cdot k_2} 
\end{eqnarray}
where ${\cal P}^0_{\mu\nu} \equiv \frac {P_{\mu\nu}}{\sqrt{3}}$ and 
${\cal P}^2_{\mu\nu} \equiv \epsilon_{\mu\nu}$.
The form factor $f(k_1^2,k_2^2)$ in Eq. \ref{extra5} is determined by
the wavefunctions $\phi(x)$ which appears in eqs. \ref{extra3} and
\ref{extra4}. 

Comparison with eqs. (7) and (10) shows that the glueball
wavefunction (\ref{extra5}) corresponds to 
$A_{0^{++}} F_1[0^{++}](k_1^2,k_2^2) = A_{2^{++}}F_1[2^{++}](k_1^2,k_2^2)$;
 the remaining form factors 
vanish.  Since this relation between $0^{++}$ and $2^{++}$ form
factors is the same as in the NRQM, both models give the same result
for the ratio of the $2^{++}$ and $0^{++}$ widths.  Thus in the limit
that the masses of the scalar and tensor states are
equal, independently of whether they are a pair of
(NRQM) $q \bar{q}$ \cite{barb,ale,zpli}
or (cavity approximation) glueball states,
\begin{equation}\label{ratio}
\frac {\Gamma(R_{2^{++}})}{\Gamma(R_{0^{++}})}=\frac 4{15}.
\end{equation}
The large mass gap between the $0^{++}$
glueball candidates $f_0(1500)$ and $f_0(1700)$, and the $2^{++}$
candidate $\xi(2230)$ (assuming $\xi(2230)$ has $J=2$) prevents
immediate application of eq. \ref{ratio}.  However since these $f_0$'s have
widths 100-150 MeV and the $\xi(2230)$ width is $\sim 20$ MeV  
(see sec. 3.3) -- either $f_0$ is compatible with (\ref{ratio}).  The
presence of $0^{++}$ $q\bar q$ states in the vicinity of these
$f_0$'s also complicates the situation (see sec. 6).

For a light $q\bar q$ system, Eq. \ref{ratio} will be modified by 
relativistic effects\cite{zpli}, which increases the ratio $4/15$ to 
around $\frac{1}{2}$.
  While there are no $R\to gg$ data available, one could relate
the  width for $R\to gg$ to that for the $R\to \gamma\gamma$ at the
tree level.  The data for $f_0(1300)\to \gamma\gamma$ and $f_2(1270)
\to \gamma \gamma$ are consistent with the result here.  However, the
relativistic effects on the loop integral $x|H(x)|^2$ remain to
be investigated.

\subsection{Virtuality Dependence of Glueball Form Factors}
\hspace*{2em}

The analysis of $J/\psi$ radiative decays in ref.\cite{cak} implicitly
assumed that the relative size of the on-shell form factors and also
their dependence on gluon virtualities are universal for $q\bar q$
mesons and glueballs. Our investigation of the previous section showed
that the first assumption may be reasonable.  

For the $<gg|R>$ form factors the situation is more complicated than
for electromagnetic form factors, because the gluon virtualities can
vary independently.  That is, the form factors here are functions of
two variables, constrained by the requirement of being even under
interchange of the gluon momenta (from Bose statistics).  We can
further constrain the form factors by power counting arguments.
Replace the variables $k_1^2$ and $k^2_2$ by $k_1\cdot k_2$ and the
dimensionless ratio $z  \equiv \frac{(k_1-k_2)\cdot (k_1+k_2)}{k_1
\cdot k_2}$.  When $R$ is an $L=0$ bound state of two constituents, the
leading large $k_1 \cdot k_2$ behavior of $F_1(k_1^2,k_2^2)$ is 
$\frac{1}{(k_1 \cdot k_2)} f(z)$.  The $F_i(k_1^2,k_2^2)$ entering $<gg|R>$
with additional factors of $k_i^{\mu}$ have correspondingly more rapid 
falloff, like $F_2$ compared to $F_1$ for the case of electromagnetic
form factors.  For $L=1$ systems one expects an additional $\mu^2/k_1
\cdot k_2$ suppression, where $\mu$ is a scale reflecting the
variation of the wavefuntion at the origin.  This scaling behavior is
manifested by the NRQM results eqs. \ref{qqbar0++} - \ref{qqbar0-+}.

Neglect of higher twist corrections to the leading form factors and
neglect of those form factors whose leading dependence falls more
rapidly, can be expected to give corrections to the $H_J(x)$'s of
order $m_R^2/M_V^2$ compared to the leading terms.  Since for our
application this is a small quantity, we neglect these corrections.  As
we saw in sec. \ref{ff}, in the NRQM $f(z) = const$. The effect of
possible corrections to constancy of $f(z)$, and overall scaling behavior 
which differs from the NRQM, is presently under study.  
 
\subsection{Higher Order Corrections and Scale Dependence}
\label{hoc} 
\hspace*{2em}
For a resonance which couples to the two-gluon intermediate state,
corrections to the above formalism involve one additional gluon loop
and thus should be of order $O(\alpha_s/4 \pi)$ in the amplitude.  For
heavy $Q \bar{Q}$ mesons the wavefunction can be treated perturbatively and
it is straightforward to make a systematic expansion in the coupling
constant\cite{barb,barbieri}.  In this context it is sensible to
distinguish between the components of the wavefunction in which the
 $Q \bar{Q}$ are in a ``color singlet'' or ``color octet''
state\cite{braaten}.  However for the light $q \bar{q}~$ mesons and
glueballs of interest, defining a perturbative expansion and the
relation between ``constituent" and ``current" partons are more subtle
and we do not undertake this here.  Suffice it to say that the concept
of ``color singlet'' versus ``color octet'' components of the
wavefunction does not have an {\it a priori} well-defined intuitive
meaning as for the heavy quark system.  The issue of composition is a
scale-dependent question, as it is for the nucleon. The
Altarelli-Parisi evolution of the parton distribution functions is a
clear illustration of this point.  In principle the same is true for
the heavy quark system, however the quark mass gives a natural scale
in that case.

Our treatment in previous sections implicitly made use of an effective
Lagrangian approach to the problem.  By working with $b_{rad}({Q \bar{Q}}_V \rightarrow \gamma +R)$, i.e.,
dividing $\Gamma({Q \bar{Q}}_V \rightarrow \gamma +R)$ by $\Gamma(Q \bar{Q}_V \rightarrow \gamma + X) $
also computed in leading order perturbation theory, one removes the
dependence on the effective strong coupling at the heavy quark
vertices.  Similarly, quoting the result in terms of $br(R \rightarrow gg)$ removes at
leading order the sensitivity to the scale dependence of the
definition of the $R$ wavefunction.  Of course, the concept of
``gluonic width'' of a $q \bar{q}~$ resonance necessarily has an intrinsic
scale dependence -- as one goes to shorter and shorter distance scale,
contributions from the parton sea invalidate simple valence intuition.
Without a careful treatment of next-to-leading order corrections, we
cannot specify the correct scale for $\alpha_s$ appearing in the
estimate of eq.(\ref{CF}) for the gluonic branching fraction of $q \bar{q}~$
mesons. For this reason, we can make only qualitative use of the
gluonic branching fractions that we extract for $q \bar{q}~$ mesons.  However
when the branching fraction of a state is found to be $\,\raisebox{-0.13cm}{$\stackrel{\textstyle>}{\textstyle\sim}$}\, 1/2$,
indicating that the state has a significant gluon component, the
sensitivity on $\alpha_s$ is a higher order correction and we {\it
can} make quantitative use of the $br(R \rightarrow gg)$ that we extract
from the data. 

It follows from the above discussion that we cannot expect a trivial
relationship between the form factors for the amplitude $<g g| R>$,
and those for $<R| \gamma \gamma>$.  In principle the latter can be
measured as a function of photon virtualities in an $e^+ e^-$
collider.  For heavy $Q \bar{Q}$ resonances such as the $\chi_c$ 
states, these amplitudes are identical except for the value of the
overall coefficient $A_{J^{PC}}$.  To obtain $<R|\gamma \gamma>$ from
$<g g| R>$, substitute $g_s \rightarrow e_Q$ and remove the color
factor.  At leading order this gives 	
\begin{equation}\label{gammafrg}
\Gamma( R \to \gamma \gamma) = \frac {9e_Q^4}{2}
(\frac{\alpha}{\alpha_s})^2 \Gamma (R \to gg).
\end{equation}
We test the validity of eq. (\ref{gammafrg}) for light $q \bar{q}~$
resonances in section \ref{gam2} by applying it to the known 
$q \bar{q}~$ states $f_2(1270)$ and $f_2(1525)$.  We then extend it to other
examples.  Insisting on the naive relation (\ref{gammafrg}) allows one
to extract an effective value of $\alpha_s$.  Doing so for several
$q \bar{q}~$ resonances gives some idea of the sensitivity of $br(R \rightarrow gg)$ to
scale.   

It might be the case that the dynamics of the form factors,
i.e., the functional dependence of the $F_i$'s on $k_1^2$ and $k_2^2$ and 
their relative normalization at the on-shell point, corresponds
more accurately than does their overall normalization, in going from
$<R|\gamma \gamma>$ to $<g g|R>$\footnote{E.g., for $q \bar{q}~$ states, the
overall normalization of $<g g|R>$ contains a factor $\alpha_s$ and is
necessarily scale dependent.}.  This could in principle be tested 
by measuring the off-shell form factors in a $\gamma \gamma$ collider
and using this dependence to predict the $H_J(x)$'s appearing in eq.
(\ref{CF}).  Once $b_{rad}({Q \bar{Q}}_V \rightarrow \gamma +R)$ is measured for both $J/\psi$ and $\Upsilon$ 
radiative decay to a given $q \bar{q}~$ meson, one can infer $H_J$ at
two values of $x$.  With several related $q \bar{q}~$ mesons of different
masses, such as $f_2(1270)$ and $f_2(1520)$, this will give a number
of points in $H_J(x)$.   

\section{Constraints From Radiative Quarkonium Decay}
\label{result}

\subsection{$J^{P}$ dependence of ${Q \bar{Q}}_V \rightarrow \gamma R_{J^P}$}
\label{Jdep}
\hspace*{2em}
The loop integral in eq. (\ref{QQgamR}) determines the function $H_J(x)$
appearing in (\ref{CF}).  For the NRQM wavefunctions these integrals
have been evaluated\cite{korner,bg} in analytical form and are
recorded in the appendix for convenience.  Readers
interested in the derivation of the analytic expressions are referred
to those papers. The relevant functions $x |H_J(x)|^2$ are shown in
Fig. 1 for $J^{PC} = 0^{++},\ 2^{++}$ and $0^{-+}$. 
The corresponding $c_R$'s in Eq. \ref{CF} are: 
\begin{equation} \label{cr}
c_R=\left \{\begin{array}{r@{\quad}l}
1 & J^{PC}=0^{-+} \\
\frac 23 & J^{PC}=0^{++} \\
\frac 52 & J^{PC}=2^{++}. \end{array}\right .
\end{equation}

In the $x$ regime of immediate interest, $x \sim 0.5 - 0.75$,  we note
from Fig. 1 that $\frac{x|H_J|^2}{30-45} \sim O(1)$. This enables us
to manipulate the CF expression, eq. (\ref{CF}), into a scaled form that 
exhibits the phenomenological implications immediately.  Specifically, 
for scalar mesons 
\begin{equation} \label{0++}
10^3  br(J/\psi \rightarrow \gamma 0^{++}) = (\frac{m}{1.5\; {\rm GeV}})  
(\frac{\Gamma_{R\rightarrow gg}}{96\; {\rm MeV}})  \frac{x|H_S(x)|^2}{35}.
\end{equation}
This is to be compared with the analogous formula for a tensor meson:
\begin{equation} \label{2++}
10^3  br(J/\psi \rightarrow \gamma 2^{++}) = (\frac{m}{1.5\; {\rm GeV}})  
(\frac{\Gamma_{R\rightarrow gg}}{26\; {\rm MeV}})  \frac{x|H_T(x)|^2}{34}.
\end{equation}
For pseudoscalars we find:
\begin{equation} \label{0-+}
10^3  br(J/\psi \rightarrow \gamma 0^{-+}) = (\frac{m}{1.5\; {\rm GeV}})  
(\frac{\Gamma_{R\rightarrow gg}}{50\; {\rm MeV}})  \frac{x|H_{PS}(x)|^2}{45}.
\end{equation}
Having scaled the expressions this way, because $\frac{x|H_J|^2}{30-45}
\sim O(1)$ in the $x$ range relevant for production of 1.3 - 2.2 GeV
states (see fig. 1), we see immediately that the magnitudes 
of the branching  ratios are driven by the denominators 96 and 26 MeV
for $0^{++}$ and $2^{++}$, and $50$ MeV for $0^{-+}$.   Thus if
a state $R_J$ is produced in $J/\psi \rightarrow \gamma X$ at
$O(10^{-3})$ then $\Gamma (R_J \rightarrow gg)$ will typically be of  
the order $100$ MeV for $ 0^{++}$, $O(25 ~ {\rm MeV})$ for $2^{++}$,
and $O(50 ~ {\rm MeV})$ for $0^{-+}$. 

This immediately shows why the $2^{++}$ $q \bar{q}~$ states are prominent: A
$2^{++}$ state with a total width of $O(100 ~\rm{MeV})$ (typical for
$2^{++}$ $q \bar{q}~$'s in this mass range\cite{cafe95,barnes95}) will be
easily visible in $J/\psi \rightarrow \gamma 2^{++}$ with branching
fraction $O(10^{-3})$, while remaining consistent with
\begin{equation} \label{f320} 
br(R[Q\bar{Q}] \rightarrow gg) = 0(\alpha^2_s) \simeq 0.1-0.2.
\end{equation}

Eqs. \ref{0++} - \ref{0-+} not only indicate which $q\bar{q}$ states
will be prominent in $J/\psi \rightarrow \gamma R$, but they also help to
resolve an old paradox concerning $0^{++}$ production.  It was
recognised early on that when the gluons in the absorbtive part of
$J/\psi \rightarrow \gamma gg$ are classified according to their
$J^{PC}$, the partial wave with $2^{++}$ was predicted to dominate.
The waves with $0^{-+}$ and $0^{++}$ were also predicted to be
significant and of comparable strength to one another \cite{bill}.
When extended to include the dispersive part\cite{cak,korner} the
$0^{++}$ was predicted to be prominent over a
considerable part of the kinematic region of interest.  States with $J
\geq 3$ were predicted to have very small rate in this process.
Experimentally, all but one of these appeared to be satisfied.  There
are clear resonant signals in $2^{++}$ and $0^{-+}$, and no
unambiguous signals have been seen with $J \geq 3$.  However no
$0^{++}$ signal had been isolated.  

>From our relations above, we see that for a $0^{++}$ to be produced at
the $10^{-3}$ level in $J/\psi$ radiative decay it must either have a
large gluonic content and width $O(100)$ MeV or, if it is a $q \bar{q}~$
meson, it must have a very large width, $\,\raisebox{-0.13cm}{$\stackrel{\textstyle>}{\textstyle\sim}$}\, 500$ MeV.  Taking this
into account, along with the following points, the puzzle of the
absence of $0^{++}$ signal has been resolved:

(i): The width of $^3P_0$ $q\bar{q}$ is predicted to be $\sim 
500$  MeV\cite{cafe95,barnes95}. Thus production at the level
$br(J/\psi \rightarrow \gamma (gg)_{0^+} \sim 10^{-3})$
is consistent with $br(R \rightarrow gg) = 0(\alpha^2_s)
\simeq 0.1-0.2$, but the $\sim 500$ MeV wide signal is smeared
over a large kinematic ($x$) range. 

(ii): The $\sim 100 ~ {\rm MeV}$ wide $f_0(1500)$ signal seen in $J/\psi
\rightarrow \gamma 4\pi$ was originally misidentified as
$0^{-+}$, but is now understood to be $0^{++}$\cite{bugg}.  

(iii): The $f_J(1710)$ which was originally believed to be $J=2$ may
contain a contribution with $J=0$\cite{bugg,pdg94}.

\subsection{Flavor Mixing and the $f_2(1270)$ and $f_2(1520)$
$q\bar{q}$ States} 
\label{ss:flav}
\hspace*{2em}
We can test this formalism by applying it first to the
the well known quarkonium states $f_2(1270)$ and $f_2(1525)$.
The above formulae have been derived for the case that the produced
meson $R(q_i\bar{q}_i)$ contains a single flavour, so we begin by
considering what changes occur for a state of mixed flavour. We shall
see that $br(J/\psi \rightarrow \gamma R)$ and $\Gamma( R \rightarrow
gg)$ depend on the mass and flavour of $R$, but in a common way such
that $br(R \rightarrow gg)$ is universal, as summarized in
eqs.\ref{0++}-\ref{0-+}. 

For a general $q \bar{q}~$ resonance
\begin{equation} \label{f321}
R \equiv cos\phi |n\bar{n} \rangle + sin \phi |s\bar{s} \rangle,
\end{equation}
where $n\bar{n} \equiv \frac{u\bar{u} + d\bar{d}}{\sqrt{2}}$.
Allowing for flavour symmetry breaking
\begin{equation} \label{f322}
\langle gg|s\bar{s}\rangle \equiv r_s^2 \langle gg|d\bar{d}\rangle .
\end{equation}
Thus
\begin{equation} \label{f323}
\langle gg|R\rangle = (\sqrt{2} cos\phi + r_s^2 sin\phi)\langle gg|d\bar{d}
\rangle,
\end{equation}
so
\begin{equation} \label{f324}
\Gamma(R \rightarrow gg) \equiv (\sqrt{2} cos\phi + r_s^2 sin\phi)^2
\Gamma(R(d\bar{d}) \rightarrow gg)
\end{equation}
and similarly
\begin{equation} \label{f325}
\Gamma(V \rightarrow \gamma R) =\Gamma(V \rightarrow \gamma R(d\bar{d})) (
\sqrt{2} cos\phi + r_s^2 sin\phi)^2.
\end{equation}

Evidently, the flavour factors cancel out in derivation of the
expressions of the previous sections and so apply immediately to
states $R$ of arbitrary flavour mixings.  We can illustrate this with
the $^3P_2$ states $f_2(1270)$ and $f_2(1520)$, for which $cos \phi
\sim 1$ and 0 respectively.  From eq. \ref{f324} we have 

\begin{equation} \label{f326}
\Gamma(f_2(n\bar{n})(1270) \rightarrow gg)/\Gamma (f_2(d\bar{d})(1270)
 \rightarrow gg) = 2
\end{equation}
and
\begin{equation} \label{f327}
\Gamma(f_2(s\bar{s})(1520) \rightarrow gg)/\Gamma (f_2(d\bar{d})(1520)
 \rightarrow gg) = r_s^2.
\end{equation}
If $q\bar{q} \rightarrow gg$ is flavour blind we expect $r_s^2 \sim 1$. 

To confront these equations with data we use the
measured radiative branching ratios\cite{pdg94}
\begin{equation} \label{f328}
10^3 \times br(J/\psi \rightarrow \gamma f_2(1270))
= 1.4 \pm 0.14
\end{equation}
and
\begin{equation} \label{f328b}
10^3 \times br(J/\psi \rightarrow \gamma f_2(1520))
= 0.63 \pm 0.1.
\end{equation}
>From eq. (\ref{2++}), we have
\begin{equation} \label{f329b}
\Gamma(1270 \rightarrow gg) = 41 \pm 7  ~ {\rm MeV}.
\end{equation}
and
\begin{equation} \label{f329}
\Gamma(1520 \rightarrow gg)
= 17 \pm 2  ~ {\rm MeV}.
\end{equation}
Combining these results with the measured widths,
\begin{equation} \label{width1}
\Gamma_{tot}(1270) = 185 \pm 20  ~ {\rm  MeV};~~ \Gamma_{tot}(1520) = 76
\pm 10  ~ {\rm MeV}, 
\end{equation}
we find 
\begin{equation} \label{f331}
br(f_2(1270) \rightarrow gg) \simeq br(f_2(1520) \rightarrow gg) = 0.22
\end{equation}
which are as expected for established $q\bar{q}$ states, see eq.(1) and ref.
\cite{cak}. Inter alia this supports the idea that glueball mixing is
not prominent in the $2^{++}$ channel at these masses\cite{cafe95}.

If the dependence on mass is weak in going from 1270 to 1520 MeV,
eqs.\ref{f326} and \ref{f327} imply
\begin{equation}
\frac{\Gamma(f_2(1270) \rightarrow gg)}{\Gamma(f_2(1520) \rightarrow
gg)} = \frac{2}{r_s^2}.
\end{equation}
Inserting the widths from eqs. \ref{f329b} and \ref{f329} we see that
$r_s \simeq 0.8 - 1$, and thus $q \bar{q}~$ flavor can be ignored to first
approximation in this analysis.  We will exploit this flavor
independence in a later section (\ref{ss:f1500}) to probe the
structure of wavefunctions for potential $q \bar{q}~$ mesons.

\subsection{Radiative Upsilon Decay}
\hspace*{2em}
No peaks are seen in the photon energy spectrum in inclusive $\Upsilon
\to \gamma X$ at a branching fraction sensitivity of about
$10^{-4}$\cite{pdg94}.  The following analysis suggests that with only
a factor of a few improvement in sensitivity, many interesting states
should become evident.  We could use data on $\Upsilon \to \gamma R$
in two ways.  Firstly, for $R \equiv c \bar{c}$, pQCD-NRQM predictions
should be reliable. Testing those predictions tests the underlying
assumption of this methodology: that pQCD provides an adequate
description of the $<Q \bar{Q}_V | \gamma gg>$ amplitude.  Secondly,
production of a given resonance in $\Upsilon \to \gamma R_J$ depends
on $H_J(x)$ at $x$ much closer to 1. This allows a more detailed
examination of the form factors, as well as probing the $x \to 1$
region where resummation of perturbation theory may be required for
the $0^{++}$ and $2^{++}$ cases.   

Let us begin by considering $\Upsilon \rightarrow \gamma \chi_c$,
where the $Q\bar{Q}$ bound states are rather well
understood\cite{korner}. The branching ratio is 
\begin{equation}\label{0171b} 
br(\Upsilon \to \gamma +R_c)=
c_J(\frac {4}{5}\frac {\alpha}{\alpha_s})
\Gamma_{R \rightarrow gg} \frac {x|H_{J}(x)|^2}{8\pi(\pi^2-9)}\frac {m}
{M^2}.
\end{equation}
where $c_J = 1(\eta_c),~ \frac{2}{3} (\chi_c^0)$ and $
\frac{5}{2}(\chi_c^2)$. In all these examples $x \equiv 1 -
m_R^2/M_{\Upsilon}^2 \sim 0.9$ so from fig. 1 we have 
$x|H|^2 = 54 (\eta_c),~ 32 (\chi_c^0),~ 37 (\chi_c^2)$. If we use
$\alpha_s(m_b) \sim 0.18$ and include the one loop corrections from
\cite{horgan}, we find
$$
br(\Upsilon \rightarrow \gamma gg) \equiv \frac{4 \alpha}{5 \alpha_s}
(1 - 2.6 \frac{\alpha_s}{\pi}) \sim 2.8 \%.
$$
Identifying $\Gamma(c \bar{c} \to gg)$ with $\Gamma(c \bar{c} \to $
light hadrons) implies 
$$
br(\Upsilon \to \gamma \chi_2) \sim br(\Upsilon \to \gamma \chi_0)
\sim 0.9 \times 10^{-5}$$
and
$$
br(\Upsilon \to \gamma \eta_c) \sim (2.3 \pm 0.9) \times 10^{-5}.
$$

Although these predicted branching ratios are small, the photons are
in a region of phase space where there is little background, so a
relatively short period of dedicated running at a B factory should be
adequate to observe these modes.  Precision data on these transitions
could both validate the pQCD analysis and give insights into higher
order effects including the role of colour octet components in the
$\chi$ wavefunctions. 

Data on $\Upsilon \to \gamma f_{0,2}(1270 - 1700)$ may also be
obtained, replacing the present upper limits $\leq
10^{-4}$\cite{pdg94}.  The kinematics here are $x \sim 0.97$. In this
region $|H_S(x)|^2$ and $|H_T(x)|^2$ are dominated by $ln(1-x)$ divergences,
and the leading order pQCD predictions become unreliable.  We urge that
studies of pQCD resummation be made in order to analyze this process and make
predictions.  Data on these processes could be used both to extract
$|H_J(x)|^2$ for phenomenological use as in sec. \ref{glpheno}, and
also permit detailed testing of pQCD resummation techniques.  The
qualitatively different behavior of the $x \to 1$ limits of $H_J(x)$
for the $0^{-+}$ and $0,2^{++}$ cases can also be exploited to this
end. 

\subsection{$1^{++}$ States}
\label{axials}
\hspace*{2em}
Before turning to our main topic of glueball candidates, we discuss
briefly the possibility of applying this formalism to $1^{++}$ mesons.
Since for them we cannot normalize the $<gg|R>$ amplitudes using the
procedure of sec. \ref{Jdep} which lead to eqs. (\ref{gam0++} -
\ref{gam2++}) it is not certain that this is possible.  However in 
the spirit of an effective lagrangian approach it might be appropriate
to consider that gluons have an effective mass, so that the amplitude
$<gg|R>$ need not vanish at the on-shell point when one of the gluons
is longitudinal.  Making the further assumption that the pQCD-NRQM
approximation of ref. \cite{barbieri} gives an adequate description of
the $1^{++}$ total width, with an adjustable overall normalization,
one can obtain a relation of the form of eq. (\ref{CF})\cite{cak}.
Substituting for $H_1(x)$ leads to the scaled formula 
\begin{equation} \label{1++}
10^3 br(J/\psi \rightarrow \gamma 1^{++}) = (\frac{m}{1.45\; {\rm GeV}})  
(\frac{\Gamma_{R\rightarrow gg}}{12\; {\rm MeV}})  \frac{x|H_1(x)|^2}{30}
\end{equation}
where in this application $\Gamma(R \rightarrow gg)$ is the total
direct coupling to gluons, not literally the coupling to two massless
gluons. 

It is interesting to apply the above relation to the $f_1(1285)$,
$f_1(1420)$ and $f_1(1530)$ states (see also ref.\cite{korner}).
Only two isoscalar mesons
can be accomodated in a quarkonium nonet and there has been
considerable discussion as to which of the three axial states is the
odd one out (and, if it exists, what its nature is). There has been
no confirmed sign
of $f_1(1530)$ in $J/\psi$ radiative decay whereas the $f_1(1420)$ and
$f_1(1285)$ are both seen. Their branching ratios are respectively\cite{pdg94}
\begin{equation}
\label{1285}
br(J/\psi \to \gamma f_1(1285)) = (0.65 \pm 0.10) \times 10^{-3}
\end{equation}
and
\begin{equation}
\label{1420}
br(J/\psi \to \gamma f_1(1420))\times br(f_1(1420) \to K\bar{K}\pi)
 = (0.83 \pm 0.15) \times 10^{-3}.
\end{equation}
Inserting these values into eq. \ref{1++} together with
$\Gamma_{tot}(f_1(1285)) = 24 \pm 3$ MeV and  $\Gamma_{tot}(f_1(1420))
= 52 \pm 4$ MeV give  
\begin{equation}
\label{gg1}
br(f_1(1285) \to gg) = 0.34 \pm 0.05
\end{equation}
and
\begin{equation}
\label{gg2}
br(f_1(1420) \to gg) = (0.19 \pm 0.04)/(br(f_1(1420) \to K\bar{K}\pi)). 
\end{equation}
If we now input $br (f_1(1420) \to K\bar{K}\pi) \sim
0.67$\cite{seiden} we find
\begin{equation}
\label{gg3}
br(f_1(1420) \to gg) \sim 0.3.
\end{equation}
Within the uncertainties of applying these ideas to low masses (e.g.$f_1(1285)$
has a low width due to phase space suppression of $KK^*$)
and the ill defined branching ratio to $K\bar{K}\pi$ for the $f_1(1420)$,
 these results are not 
inconsistent with the accepted $q \bar{q}~$ interpretation of the
$f_1(1285)$ and support also the quarkonium interpretation of $f_1(1420)$
(unless the $br(K\bar{K}\pi)$ should turn out to be much overestimated).

The ratio
\begin{equation}
\label{axialratio}
\frac{br(J/\psi \to \gamma f_1(1420))}{br(J/\psi \to \gamma f_1(1285))} \sim 1.9
\end{equation}
is consistent with the quarkonium mixing arising from a quadratic mass
formula for the axial nonet\cite{seiden}
\begin{equation}
\label{axialmix}
f_1(1285) = 0.94 |n \bar{n} \rangle - 0.35 |s \bar{s} \rangle
\equiv 0.57 |1 \rangle + 0.83 | 8 \rangle 
\end{equation}
\begin{equation}
\label{axialmix2}
f_1(1420) = 0.35 |n \bar{n} \rangle + 0.94 |s \bar{s} \rangle.
\equiv 0.83 | 1 \rangle - 0.57 | 8 \rangle
\end{equation}

Recent data from BES\cite{besiota} have large error bars but
are consistent with the older data
for $f_1(1420)$;
they obtain $
br(J/\psi \to \gamma f_1(1420))\times br(f_1(1420) \to K\bar{K}\pi)
 = (0.76 + 0.46 -0.18) \times 10^{-3}$, $\Gamma = 59 \pm 5$ MeV. They
do not see any $f_1(1285)$ but this may not be surprising 
since they are looking in the $K\bar{K}\pi$ mode. They also
report a signal $f_1(1497)$, $\Gamma=
44 \pm 7$ MeV and
$
br(J/\psi \to \gamma f_1(1497))\times br(f_1(1497) \to K\bar{K}\pi)
 = (0.52 \pm 0.23) \times 10^{-3}$. This state's parameters are
also consistent with those expected for a quarkonium as long as $br(f_1(1497) 
\to K\bar{K}\pi) \geq 0.5$.

The axial mesons are currently an enigma. There are three candidates
where the quark model would require only two. The lattice predicts the 
lightest $1^{++}$ glueball to be at $\sim 4$ GeV\cite{ukqcd}.
It is noticeable that no single experiment sees all three and one
should be cautious as to whether there are indeed three genuine
states. We urge that BES, in particular, seek three $f_1$ signals or
place limits against them in order to help clarify the above analysis.
In any event, more detailed theoretical work, specifically formalizing
the effective lagrangian treatment of the problem, is warranted in
order  to relate to the production of axial mesons in $\gamma
\gamma^*$ and to provide a more solid foundation to the theoretical
analysis after the experimental situation becomes clear.

\section{Glueball Candidates}
\label{glpheno}
\subsection{$f_0(1500)$}
\label{ss:f1500}
\hspace*{2em}
We look first at the established scalar meson, $f_0(1500)$. As we shall see
below, at present there are discrepancies between the values of
$br(J/\psi \to \gamma f_0(1500))$ as determined from various experimental
analyses.

The analysis of ref. \cite{bugg} gives $ br(J/\psi \rightarrow \gamma f_0
(1500) \rightarrow \gamma \sigma \sigma) = (5.7 \pm 0.8) \times
10^{-4}$ with an overall $\pm 15\%$ normalisation uncertainty. The
analysis of ref.\cite{bugg3,buggpriv} implies that the $\sigma \sigma$
mode is at most $50 \%$ branching ratio and so we infer
$$
br(J/\psi \rightarrow \gamma f_0(1500)) \ge (1.15 \pm 0.15) \times
10^{-3} \pm 15\%. 
$$
In this case, with $\Gamma_{\rm tot}(f_0(1500))$ = $120\pm20$ MeV, if
we add errors in quadrature and use the central value, 
eq. \ref{0++} implies that  

\begin{equation} 
\label{f400}
br(f_0(1500) \rightarrow gg) \ge 0.9 \pm 0.2.
\end{equation} 
This is significantly larger than the $O(\alpha_s^2)$ which would be
expected for a pure $q\bar{q}$ system, and supports this state
as a glueball candidate.

On the other hand BES has recently reported\cite{landua} $br(J/\psi \to
\gamma f_0(1500) \to \gamma \pi^o \pi^o) = 3-5 \times 10^{-5}$.
Landua\cite{landua} combines this with the Crystal Barrel data on
$br(f_0(1500) \to \pi\pi)$ to get 
$$
br(J/\psi \to \gamma f_0(1500)) = (0.4-0.6)\times 10^{-3}.
$$
Thus via eq. (\ref{0++}), 
\begin{equation} \label{f400b}
br(f_0(1500) \rightarrow gg) = 0.3-0.5.
\end{equation} 

The interpretation of this state cannot be settled until the experimental
situation clarifies. An order of magnitude increase in statistics for
$J/\psi \rightarrow \gamma \pi \pi \pi \pi$ will enable extension
of the analysis shown in fig.2 of ref.\cite{bugg}, and $J/\psi \to \gamma
\pi \pi$ likewise needs to be improved.  The neutral channel $J/\psi
\rightarrow \gamma \pi^o \pi^o \pi^o \pi^o$ is particularly
advantageous here as it is free from $\rho \rho$ contamination
and so can help to improve the quantification of $br(f_0(1500)
\rightarrow \sigma \sigma)$.  These should be high priorities at a
$\tau$-Charm Factory.   

We shall return to the interpretation of the $f_0(1500)$ in section 6.
  
\subsection{$f_J(1710)$}
\label{ss:f1710}
\hspace*{2em}
The case of $f_{J}(1710)$ is particularly interesting and the
conclusions depend critically on whether $J=0$\cite{bugg,ll92} or
$J=2$\cite{pdg94,pdg92,wa7689}.  It has been observed most clearly
in radiative $J/\psi$ 
decay in the $K \bar{K}$ mode\cite{pdg92}, with evidence also
in the $4 \pi$ mode\cite{bugg}. Recently BES has reported seeing
both $J=0$ and $J=2$ states in this region.  We discuss the various
measurements in turn. 

In the $K \bar{K}$ channel, $br(J/\psi \rightarrow \gamma
f_J(1710) \rightarrow \gamma K\bar{K}) = (0.97 \pm 0.12)  \times
10^{-3}$\cite{pdg94}.  Assuming first that $f_J(1710)$ is a single
state with $J=2$, we use eq. (\ref{2++}): 
$$
\label{2++rr}
10^3  br(J/\psi \rightarrow \gamma f_{2}(1710)) = (\frac{m}{1.5\;
{\rm GeV}}) (\frac{\Gamma_{R\rightarrow gg}}{26\;   {\rm MeV}})
\frac{x|H_T(x)|^2}{34},
$$
which implies
\begin{equation} \label{2++r2}
\Gamma(f_2(1710) \rightarrow gg) = 
\frac{(22 \pm 3) ~ {\rm MeV} }{br(f_2(1710)
\rightarrow K\bar{K})}.
\end{equation}
No comparable signal has been seen in any other
channel in $J/\psi \to \gamma f_2(1710) \to \gamma X$ and
it would thus appear that $K\bar{K}$ is a major mode of any $J=2$
object in $J/\psi$ radiative decays 
(the listing of decay channels for $J/\psi \to \gamma f_J(1710) \to
\gamma K\bar{K}$ in ref.\cite{pdg94}
suggests that this mode is greater than
$\sim 50\%$ of the $K\bar{K} + \pi \pi + \eta \eta$ channels together).
With $\Gamma_{tot}(f_J(1710)) \sim 150$ MeV, eq. \ref{2++r2} and 
$br(f_2(1710) \rightarrow K \bar{K}) \ge 0.5$ imply
\begin{equation} \label{2++r3}
br(f_2(1710) \rightarrow gg) \,\raisebox{-0.13cm}{$\stackrel{\textstyle<}{\textstyle\sim}$}\, 30\%,
\end{equation}
which would be consistent with this state being a $q\bar{q}$.  

By contrast, if $f_J(1710)$ is a single state with $J=0$, 
we use eq. (\ref{0++}):
$$
\label{0++r}
10^3  br(J/\psi \rightarrow \gamma 0^{++}) = (\frac{m}{1.5\; {\rm GeV}})  
(\frac{\Gamma_{R\rightarrow gg}}{96\;   {\rm MeV}})
\frac{x|H_S(x)|^2}{35},
$$
which implies
\begin{equation} \label{0++r2}
\Gamma(f_0(1710) \rightarrow
gg) = \frac{(78 \pm 10)~  {\rm MeV}}{br(f_0(1710)
\rightarrow K\bar{K})}
\end{equation}
and hence
\begin{equation} \label{0++r3}
br(f_0(1710) \rightarrow gg) \ge 0.52 \pm 0.07,
\end{equation}
in accord with fig. 14 of ref.\cite{cak}.  In this case 
the $f_0(1710)$ would be a strong candidate for a
scalar glueball.  Knowing the spin and $K \bar{K}$ branching fraction
of $f_J(1710)$ is of great importance for a more detailed quantitative
understanding of the composition of this state.   

These questions have become central in view of new data from
BEPC\cite{bepc96} which, for the first time, separates a $J=2$ and
$J=0$ signal from the ``$\theta(1710)$" region. They find an
$f_2(1696)$ with $\Gamma =103 \pm 18$ MeV and $ br(J/\psi \to \gamma f_2) 
\times br( f_2 \to K^+K^-) = 2.5 \pm 0.4(10^{-4})$.  They also find an
$f_0(1780)$ with $\Gamma =85 \pm 25$ MeV and $ br(J/\psi \to \gamma f_0) 
\times br(f_0 \to K^+K^-)= 0.8 \pm 0.1(10^{-4})$. These signals are weak,
$O(10^{-4})$, in contrast to the $O(10^{-3})$ reported in the earlier
literature cited above. The BEPC $J=2$ state is consistent with a
(radial excited) $q\bar{q}$.  Their $J=0$ state strength appears
too feeble for a glueball, unless $K^+K^-$ is a minor decay mode. 

If the BEPC data are definitive, then the possibility that
$br(f_0(1780) \to K^+K^-)$ is small merits investigation. In this
context we note that ref. \cite{bugg} analyzes $J/\psi \to \gamma
4\pi$ and finds a signal at about 1750 MeV consistent with $0^{++}$,
although $2^{++}$ is not absolutely excluded.  If interpreted as
$f_0$, the width is $ \Gamma = 160$ MeV and branching fraction in
$J/\psi \to \gamma f_0 \to \gamma 4\pi$ is $ (0.9 \pm 0.13) \times
10^{-3}$\cite{bugg}. Thus the analysis of ref \cite{bugg} indicates
that there is a scalar signal in $J/\psi \to \gamma 4\pi$ at strength
characteristic of gluonic states.  We urge that BEPC investigate the
$4\pi$ channel to see if their scalar state is visible at a level
consistent with the above analysis of ref.\cite{bugg}.

A possible explanation of the observations, if both $f_0(1500)$ and
$f_0(``1710'')$ are produced at the $10^{-3}$ level in $J/\psi$
radiative decay, is that both of them contain both $q \bar{q}~$ and $gg$
components \cite{cafe95,ct96,genovese,wein96}.  Better data,
especially on the decay branching fractions and production in
radiative $J/\psi$ decay is crucial for resolving this question.  We
will consider complementary tests for this, through $\gamma \gamma$
production, in section \ref{gam2}.  We examine mixing phenomenology in
section 6.  

\subsection{$\xi(2230)$ Tensor Glueball candidate}
\label{ss:xi2330}
\hspace*{2em}  
The appearance of a narrow state $\xi(2230)$ in $J/\psi
\rightarrow \gamma \pi^+ \pi^-; \gamma K^+ K^-;$ $ \gamma K_s^o K_s^o;$  
$\gamma p \bar{p}$ has created considerable interest\cite{beijing}. In
each of these channels the branching ratios are typically $br(J/\psi
\rightarrow \gamma \xi) \times br(\xi \rightarrow X\bar{X}) \sim 3
\times 10^{-5}$ for each of the channels where $X \equiv \pi, K^+$ or
$K_s^o$, and $\sim 1.5 \times 10^{-5}$ for $p\bar{p}$. In all channels
the signal is consistent with $\Gamma_{tot} \sim 20 ~ {\rm MeV}$.
After allowing for associated neutral modes such as $\pi^o \pi^o$,
$\eta \eta$ and $n\bar{n}$ by isospin, this gives 
\begin{equation} \label{f500}
br(J/\psi \rightarrow \gamma \xi) \geq 0.1 \times 10^{-3}.
\end{equation}
When combined with our formulae eqs. \ref{0++} and \ref{2++} this
implies that $br(\xi \rightarrow gg) \geq 0.4$ for $J=0$
and $\geq 0.15$ for $J=2$.  However eq. (\ref{f500}) is likely to be a
gross underestimate because $\rho \rho$, $ \omega \omega$ and multibody
channels were not included.  Indeed, the absence of a signal in 
PS185 at CERN\cite{ps185}, suggests that two body final states
constitute no more than $\sim 10 \%$ of the total.  In view of the
uncertainties in the measured quantities this gives $br(\xi
\rightarrow gg)$ consistent with unity for $J=2$; for 
$J=0$, it unacceptably exceeds the unitarity bound.   

Thus we suggest that if evidence for the $\xi(2230)$ survives
increases in statistics, the case $J=2$ would be consistent with
$\xi(2230)$ being a tensor glueball.  Such a result would have
significant implications for the emergence of a glueball spectroscopy
in accord with lattice QCD. It would also raise tantalising questions
about the $0^{-+}$ sector, where lattice finds a glueball mass $\geq ~
2~{\rm GeV}$.  This is interesting in view of the appearance of a clear
$0^{-+}$ signal in the $1450$ MeV region, which would then be
difficult to reconcile with being the $0^{-+}$ glueball.  We now turn
to this question.

\subsection{$0^{-+}$ signals in $J/\psi \rightarrow \gamma
R$}
\hspace*{2em}
The branching ratio for $(Q\bar Q)_V\to \gamma R$ in terms of the
total gluonic decay width of the pseudoscalar state is given by eqs.
1 and \ref{cr}
\begin{equation}
b_{rad}((Q\bar Q)_V\to \gamma +R)=
\Gamma_{R \rightarrow gg} \frac {x|H_{PS}(x)|^2}{8\pi(\pi^2-9)}\frac {m}
{M^2}.\nonumber
\end{equation}
As noted in eq. (\ref{0-+}), the above formula may be scaled as
follows 
\begin{equation}
\label{0-+rate}
10^3  br(J/\psi \rightarrow \gamma 0^{-+}) = (\frac{m}{1.5\; {\rm GeV}})  
(\frac{\Gamma_{R\rightarrow gg}}{50\;   {\rm MeV}})  \frac{x|H_{PS}(x)|^2}{45}.
\end{equation}
This subsumes figs 2,3,5,6, 9 and 10 of ref.\cite{cak}.  (Note 
that the dashed curve in fig 8 of ref. \cite{cak} corresponds
to the above; the figure caption has typographical errors).

As a consistency test of this methodology for $0^{-+}$ states, we
consider the production of the radial $q \bar{q}~$ anticipated in this
mass region\footnote{Note that the model cannot safely be applied to
the $\eta'$, whose total width is ``accidentally" strongly suppressed
because (apart from $\eta \pi \pi$, which is suppressed by three body
phase space) only electromagnetic decays are non-negligible}.   The
state $\eta(1295)$ ($\Gamma = 53 \pm 6 $ MeV, with dominant decay into
$\eta \pi \pi$) is a candidate on the grounds of mass and
width\cite{barnes95}. The DM2 collaboration \cite{dm2} may have
evidence for the $\eta(1295)$ in their $J/\psi \rightarrow \gamma \eta
\pi \pi$ data, which contains a peak in $\eta \pi \pi$ with the parameters
$J^{PC} = 0^{-+}$, $m=1265$, $\Gamma = 44 \pm 20 {\rm MeV},~ br =
(0.26 \pm 0.06) \times 10^{-3}$. If this is the $\eta(1295)$, the
scaled formula  (\ref{0-+}) then implies that $br(\eta(1295)
\rightarrow gg) \sim 0.25$ if $\eta \pi \pi$ is the dominant decay
mode.  This is consistent with a $0^{-+}(q\bar{q})$ because $\eta \pi
\pi$ dominance is expected\cite{barnes95}. 

At slightly higher mass, the $\eta(1440)$\cite{pdg94} is more
prominently produced. Historically this was seen in $J/\psi \to \gamma
K\bar{K}\pi$ with $br = (4.3 \pm 1.7) \times 10^{-3}$ \cite{iotaold}.
The prominence of this state caused it to be identified as a potential
glueball\cite{ishikawa}.  Subsequently it was realised that there are
two states contained within this structure\cite{pdg94} whose
individual production rates were smaller than the earlier, apparently
large value.  This development, together with improved lattice QCD
estimates of the $0^{-+}$ glueball mass which place it above $2$ GeV
rather than in the $1.4$ GeV region, caused the glueball
interpretation to fall from favour.  As noted in ref. \cite{cak}, such a
large production rate as originally reported\cite{iotaold} 
would seriously oversaturate $br(\eta(1440) \to gg)$, but subsequent
separation of the signal into two resonances results in physically
acceptable values for the individual $gg$ widths.

The first analyses\cite{bai90,dm2} indicating the existence of
additional structure in the $\eta(1440)$ region were, however, not
in agreement.  Recent data in $p\bar{p} \to \eta(1440) + \cdots $ help us to
identify the problematic measurement and to propose a consistent
picture that experiments should now pursue. We shall suggest that
there are two states, $\eta_L$ and $\eta_H$ (for ``Low" and ``High"
mass respectively), where $\eta_L$ has significant coupling to glue
while $\eta_H$ is dominantly the $s\bar{s}$ member of the nonet, mixed
with glue.  Before giving the theoretical analysis, we survey the
evidence from various experiments for $\eta_L(1410) \to \eta \pi \pi$
and $K\bar{K}\pi$, $\Gamma_{tot} \sim 50$ MeV and for $\eta_H(1480)
\to  K^*K$, $\Gamma_{tot} \sim 100$ MeV.  

Obelix\cite{obelixE} sees two states in $p \bar{p} \to \pi \pi \eta_{L,H} 
\to \pi \pi( K\bar{K} \pi)$ with the properties 
\begin{eqnarray}\label{obeliota}
\eta_L(1416 \pm 2) \to K\bar{K}\pi; \Gamma_{tot} = 50 \pm 4~{\rm MeV} \\
\eta_H(1460 \pm 10) \to K^*\bar{K}; \Gamma_{tot} = 105 \pm 15 ~{\rm MeV}.
\end{eqnarray}
These values agree with the central values for the sighting
by MarkIII\cite{bai90} in $J/\psi$ radiative decay. Combining errors
in quadrature MarkIII finds
\begin{eqnarray}
\eta_L(1416 \pm 10) \to a_0 \pi \to K\bar{K}\pi; 
\Gamma_{tot} = 54^{ +40}_{ -30} ~{\rm MeV} \\
\eta_H(1490 \pm 18) \to K^*\bar{K}; \Gamma_{tot} = 91 \pm 68 ~{\rm MeV}.
\end{eqnarray}
Further evidence for the low mass state, in the decay channel $\eta
\pi \pi$, comes from MarkIII\cite{burchell} who find $\eta(1400 \pm
6)$, $\Gamma_{tot} = 47 \pm 13$ MeV; from DM2\cite{dm2} who find
$\eta(1398 \pm 6)$, $\Gamma_{tot} = 53 \pm 11$ MeV; and the Crystal
Barrel Collaboration\cite{cbiota}. The latter see $p\bar{p} \to \pi
\eta(1410 \pm 3) \to \pi (\eta \pi \pi)$ with significant contribution in
the glue favoured partial wave $\eta \sigma$\cite{cafe95}. 
Their value $\Gamma_{tot} = 86 \pm 10$ MeV is however substantially
larger than the $\Gamma_{tot} \approx 50$ MeV found by the other
experiments\cite{obelixE,dm2,burchell}.  Possibly the differing
production mechanisms of these experiments affects the apparent width
of the resonance due to differing interference effects. 

We now compute the production rate for these states in $J/\psi$
radiative decay.  For the $\eta_L$, MarkIII\cite{bai90}
see the decay mode $a_0 \pi \to K\bar{K} \pi$ with
\begin{eqnarray}\label{mk3L}
br(J/\psi \to \gamma \eta_L(1410) \to \gamma K\bar{K}\pi) =
(0.66^{ +0.29}_{ - 0.22}) \times 10^{-3},
\end{eqnarray}
while a clear signal is found also in the $a_0 \pi \to \eta \pi \pi$
channel by ref.\cite{burchell}:
\begin{eqnarray}
br(J/\psi \to \gamma \eta_L(1410) \to \gamma \eta \pi \pi) =
(0.34 \pm 0.08) \times 10^{-3}.
\end{eqnarray}
These two channels are expected to dominate the decays.  Adding them
together and inserting into eq. (\ref{0-+rate}) implies
\begin{equation}
\label{1400gg}
\Gamma(\eta(1410) \to gg) = (54 \pm 13)~ {\rm MeV}.
\end{equation}
Combining the various width measurements, adding errors in quadrature,
gives $\Gamma_{tot} = 54.2 \pm 3.4$ MeV, and hence
\begin{equation}
br(\eta(1410) \to gg) = 0.9 \pm 0.2.
\end{equation}
For comparison, using just the larger width from Crystal
Barrel\cite{cbiota} would give $br(\eta(1410) \to gg) = 0.65 \pm 0.2$.
The data clearly indicate a strong coupling to gluons which argues
against $\eta_L(1410)$ being pure $q\bar{q}$. Rather, it couples like 
a glueball, perhaps mixed with the nearby $q\bar{q}$ nonet.

Now we consider the $\eta_H(1480)$. This state decays into $K^*K$ and
is not seen in $\eta \pi \pi$.\footnote{For this reason the DM2
analysis\cite{dm2} indicating $\eta(1460) \to a_0 \pi \to K\bar{K}\pi$
is suspect, since a genuine resonance decaying to $a_0 \pi$ should
also show up in $a_0 \pi \to \eta \pi \pi$.  Instead, in the $\eta \pi
\pi$ channel only the $\eta(1410)$ is seen.  Eliminating this DM2
state produces a harmonious picture given the remaining observations.
We thank A.Kirk for discussions of this point.}  Combining errors in
quadrature, as above, MarkIII finds\cite{bai90} 
\begin{equation}
\label{mk3kk}
br((J/\psi \to \gamma \eta(1490) \to \gamma K^* K) = 1.03^{ +0.33}_{
-0.26} (10^{-3})
\end{equation}
whereby with $\Gamma_{tot} =100 \pm 20$ MeV\cite{obelixE}, eq.
(\ref{0-+rate}) implies
\begin{equation}
\label{etaHrate}
br(\eta(1490) \to gg) = (0.5 \pm 0.2)/br(K^*K).
\end{equation}

Ref\cite{barnes95} anticipates that the radially excited $\eta^{s\bar{s}}$
should have a total width of up to 100 MeV, dominated by the channel $K^*K$.
The above result, eq. (\ref{etaHrate}), is compatible with such a state.
Ref.\cite{barnes95} also finds that $\eta^{n\bar{n}}$ has suppressed
width, decaying into $\eta \pi \pi$ through $a_0 \pi$; this is compatible with
the results on $\eta(1295)$ above. Thus a tentative interpretation of the
pseudoscalar states is as follows:
\begin{eqnarray}
\label{pseudo}
\eta(1295) \sim \eta^{n\bar{n}}; ~br_{gg} \sim O(\alpha_s^2) \sim 0.25 
\nonumber \\
\eta_L(1410) \sim G(+{q\bar{q}}); ~br_{gg} \sim 1 \nonumber \\
\eta_H(1480) \sim \eta^{s\bar{s}}(+G); ~br_{gg} \sim 0.5 
\end{eqnarray}
These conclusions can be sharpened if the widths and decays from
Crystal Barrel and Obelix converge and if $J/\psi \to \gamma 0^{-+}$
is pursued further. We note also a new measurement from
BEPC\cite{besiota} which sees only a single state with
a mass of $1467 \pm 3$ MeV, $\Gamma = 89 \pm 6$ MeV and $br(\psi \to
\gamma \eta(1467) \to \gamma K\bar{K}\pi) = 1.86 \pm 0.10 \pm 0.4
(10^{-3})$.  Since $br(\eta(1467) \rightarrow   K\bar{K}\pi) \le
1$, this gives $br(\eta(1467) \rightarrow gg) \ge 1.1 \pm 0.2$.
We urge that BEPC continue to investigate this state with a view to
separating two signals: $\eta_H \to K^*K \to K\bar{K} \pi$ and
$\eta_L \to a_0 \pi \to K\bar{K} \pi$.

The experimental data on $0^{-+}$ production in radiative $J/\psi$
decays in this mass region need clarification before strong
conclusions can be drawn, but if the existence of two states in
the $1400 -1500$ MeV range, and their relative production 
(one or both much more strongly produced than $\eta(1295)$) is
confirmed, we have a serious challenge to theoretical expectations.
The experiments would appear to be telling us that the lightest
pseudoscalar glueball is much lighter than predicted in quenched
lattice QCD, ($2.16 \pm 0.27$ GeV\cite{ct96}).  In view of the
apparent possible success (within uncertainties noted above) of the
lattice QCD predictions for the $0^{++}$ and $2^{++}$ glueball masses,
such a discrepancy between lattice QCD and nature would be of great
interest.  We note that the mass and properties of the $\eta(1410)$ are
consistent with predictions for a gluino-gluino bound
state\cite{gluino,cak}, possibly mixed with nearby pseudoscalar $q
\bar{q}$ states.  If nature were supersymmetric and SUSY breaking did
not violate $R$-invariance, the gluino mass would be $O(100)$
MeV\cite{f:99,f:110}.  In that case the $0^{++}$ glueball would be in
an approximate supermultiplet with the pseudoscalar gluino-gluino
($\tilde{g} \tilde{g}$) and spin-1/2 gluon-gluino bound states.  This 
would lead to an ``extra'' isosinglet pseudoscalar in the spectrum,
with mass around 1 1/2 GeV.  Decay of such a $\tilde{g} \tilde{g}$
system would necessarily go through gluons, since its direct couplings
to quarks would be suppressed by heavy squark masses and hadrons
containing a single gluino would be too massive to be pair produced by
the $\tilde{g} \tilde{g}$.  Thus $br(\tilde{g}\tilde{g} \to gg) \sim 1$.

Improving the data on these states would provide important
constraints.  It is now a clear challenge for experiment to separate
and quantify these signals. 

\section{Constraints on Glueballs from $\gamma\gamma$ Widths}
\label{gam2}
\subsection{$R \to \gamma \gamma$ and $J/\psi \to \gamma R$}
\hspace*{2em}
If a state $R_J$ is a glueball (or light gluinoball), it will occur in $\psi
\to \gamma R_J$ as a singleton and be strongly suppressed in $R_J \to
\gamma \gamma$.  By contrast, if $R_J$ is an $I=0$ member of a
$q\bar{q}$ nonet there will be two orthogonal states in the singlet -
octet flavour basis available for production both in $J/\psi \to \gamma R_J$ 
and $R_J \to \gamma \gamma$. Flavour $1-8$ mixing angles may suppress
one or the other of the pair in either $R_J \to \gamma \gamma$ 
or in $J/\psi \to \gamma R_J$ but there are strong correlations between
the two processes so that a comparison of the two processes can help
to distinguish glueball from $q\bar{q}$.  In particular, if a
$q\bar{q}$ state is flavour ``favoured" in $J/\psi \to \gamma q\bar{q}$,
so that it is prominent and superficially somewhat ``glueball - like",
it will also be flavour favoured in $\gamma \gamma \to R(q\bar{q})$
(see below) in dramatic contrast to a glueball.

For heavy $Q\bar{Q}$ resonances such as the $\chi_c$ states, the
amplitudes $<g g| R>$, which enter the computation of $b_{rad}({Q \bar{Q}}_V \rightarrow \gamma +R)$, and
$<R| \gamma \gamma>$, which in principle can be measured as a function
of photon virtualities, are identical except for the value of the 
overall coefficient $A_{J^{PC}}$.  The relative rates (see eq.
(\ref{gammafrg} and sec. \ref{hoc}) would be (in leading order)

$$\Gamma(R \to \gamma \gamma) = \frac{9 e_Q^4}{2} (\frac{\alpha}{\alpha_s})^2
\Gamma(R \to gg)$$
where $e_Q$ is the relevant quark charge.  Bearing in mind the
limitations to use of this relation for light $q \bar{q}~$ mesons discussed
in sec. \ref{hoc}, (see also ref.\cite{chan}) we shall tentatively
adopt this relation and test it against the known $f_2(1270;1525)$.
Finding it to be qualitatively reasonable, we apply it in section
\ref{3mixing} to $f_0(1370,1500,1710)$, allowing for mixing between $n 
\bar{n}$, $s \bar{s}$, and $gg$. 

\subsection{Orthogonal $q \bar{q}~$ mesons coupling to $\gamma \gamma$ and $gg$}
\hspace*{2em}
Define
\begin{equation} \label{f5101}
R_I\equiv cos\theta |1 \rangle + sin \theta |8 \rangle
\end{equation}
\begin{equation} \label{f5102}
R_{II} \equiv cos\theta |8 \rangle - sin \theta |1 \rangle,
\end{equation}
where $|1,8\rangle$ denote the $SU(3)$ flavor $q \bar{q}~$ states.
(This is a more natural basis for what follows than the ideal flavour basis
used in section \ref{ss:flav}).  Then in terms of the intrinsic rates
for a single $q\bar{q}$ flavour ($u\bar{u}; d\bar{d}$ or $s\bar{s}$ assumed
to be of equal strength)\footnote{The factor 3 in this equation reflects 
the $1/\sqrt{3}$ projection of each of the three $q \bar{q}$ flavors
in the flavor singlet state.} 
\begin{equation}
\label{f5103}
\Gamma(R_I \to gg) = \Gamma(q\bar{q} \to gg) \times 3 ~ cos^2\theta 
\end{equation}
and
\begin{equation}
\label{f5104}
\Gamma(R_{II} \to gg) = \Gamma(q\bar{q} \to gg) \times 3~ sin^2\theta ,
\end{equation}
while the $\gamma \gamma$ widths are in a different proportion.  
Defining $\Gamma(q\bar{q} \to \gamma \gamma) $ to be the
$\gamma\gamma$ width for quarks of unit electric charge:
\begin{equation}
\label{f5205}
\Gamma(R_I \to \gamma \gamma) = \Gamma(q\bar{q} \to \gamma \gamma) 
(cos \theta \frac{2}{3\sqrt{3}} + sin \theta \frac{1}{3 \sqrt{6}})^2
\end{equation}
and
\begin{equation}
\label{f5206}
\Gamma(R_{II} \to \gamma \gamma) = \Gamma(q\bar{q} \to \gamma \gamma) 
(-sin \theta \frac{2}{3\sqrt{3}} + cos \theta \frac{1}{ 3 \sqrt{6}})^2.
\end{equation}
In a form that shows the relation to the $gg$ widths,
\begin{equation}
\label{f5103b}
\Gamma(R_I \to \gamma \gamma) = \Gamma(q\bar{q} \to \gamma \gamma)
 \times \frac{1}{6} cos^2(\theta - \tau) 
\end{equation}
and
\begin{equation}
\label{f5104b}
\Gamma(R_{II} \to \gamma \gamma) = \Gamma(q\bar{q} \to \gamma \gamma)
 \times \frac{1}{6} sin^2 (\theta - \tau),
\end{equation}
where $\tau \equiv tan^{-1} \frac{1}{2 \sqrt{2}} \sim 19.5^o$.

It is clearly possible for an individual $q \bar{q}~$ to decouple ``accidentally"
in $gg$ or $\gamma \gamma$ if $\theta \sim 0$ or $19.5^o$.
However for the orthogonal {\bf{\it system}} we have the sum rule
\begin{equation}
\label{sum1}
\Gamma (R_I \to \gamma \gamma) + \Gamma (R_{II} \to \gamma \gamma) =
\frac {1}{6} \Gamma(q\bar{q} \to \gamma 
\gamma)
\end{equation}
and
\begin{equation}
\label{sum2}
\Gamma (R_I \to gg) + \Gamma (R_{II} \to gg) =  3 \Gamma(q\bar{q} \to gg).
\end{equation}
Thus using eq. \ref{gammafrg} and including the next order QCD corrections,
\begin{equation}
\label{f5211}
\frac{\Gamma (R_I \to \gamma \gamma) + \Gamma (R_{II} \to \gamma
\gamma) }{\Gamma (R_I \to gg) + \Gamma (R_{II} \to gg) }
= \frac{\alpha^2}{4 \alpha_s^2} \times (1 + c\frac{\alpha_s}{\pi})^{-1}
\end{equation}
where $c \sim -0.4$ for tensors and $\sim 8.6$ for
scalars\cite{barbieri,rosner}.

In the case of tensors the input data are
\begin{equation}\label{f52000}
\Gamma (f_2(1270) + f_2(1525)) \to \gamma \gamma) = 3.0 \pm 0.4 ~ {\rm keV}
\end{equation}
and, from our analysis in section \ref{ss:flav},
\begin{equation}\label{f52001}
\Gamma (f_2(1270) + f_2(1525)) \to gg)= 58 \pm 8  ~ {\rm MeV}.
\end{equation}
With these widths, eq. \ref{f5211} gives $\alpha_s^{eff} \sim 0.48 \pm
0.05$, not unreasonable for this mass region\cite{webber94}.  

Considering the ratio of the $\gamma \gamma$ and $gg$ widths of the
entire orthogonal system allowed us to extract $\alpha_s^{eff}$, with
little sensitivity to $\theta$.  We can instead employ ratios of the
$R_I$ and $R_{II}$ $\gamma \gamma$ and $gg$ widths to extract $\theta$
with little sensitivity to $\alpha_s^{eff}$.  For an orthogonal $q
\bar{q}$ pair, eqs. \ref{f5205}-\ref{f5104b} imply 
\begin{equation}
\label{f5601}
\frac{\Gamma(R_I \to gg)}{\Gamma(R_{II} \to gg)} = \frac {1}{tan^2 \theta}
\end{equation}
while
\begin{equation}
\label{f5603}
\frac{\Gamma(R_I \to \gamma \gamma)}{\Gamma(R_{II} \to \gamma \gamma)}
 = \frac {1}{tan^2 ( \theta - 19.5^o)}.
\end{equation}

Note that the two sets of equations will not give the same value of
$\theta$ if our procedure is not valid.  As a consistency check, we
determine $\theta$ both ways for the $f_2$ states. The $J/\psi \to
\gamma R$ data gave us 
\begin{equation}
\label{f5604}
\frac{\Gamma(f_2(1270) \to gg)}{\Gamma(f_2(1525) \to gg)}
 = \frac {41 \pm 7   {\rm MeV}}{17 \pm 2  {\rm MeV}} \rightarrow 
\theta = (33 \pm 2)^o,
\end{equation}
while $\gamma \gamma$ data give
\begin{equation}
\label{f5605}
\frac{\Gamma(f_2(1270) \to \gamma \gamma)}{\Gamma(f_2(1525) \to \gamma \gamma)}
 = (26.2 \pm 2.8) \pm 26\% \rightarrow \theta = (30.5 \pm 2)^o.
\end{equation}
 
The consistency of these results encourages us to apply the ideas to
scalar mesons. However, the presence of possibly three scalar
states in close proximity, $f_0(1370;1500)$ and $f_{J=0?}(1710)$,
and in the vicinity of the lattice scalar glueball, suggests that
mixing involving both $q \bar{q}$ and $gg$ will be essential.  We
shall now consider this situation. 

\subsection{$q\bar{q}$ nonet and glueball coupling to $\gamma \gamma$ and $gg$}
\hspace*{2em} 
If the $q\bar{q}$ nonet, $R_{I,II}$ is in the vicinty of a glueball,
$G$, the above analysis requires generalisation. Three isoscalars arise.
With $R_{I,II}$ as above, the mixed states may be written

\begin{eqnarray}
\Psi_3 & = & cos \beta |R_{II}\rangle - sin \beta |G \rangle \nonumber \\
\Psi_2 & = & cos \gamma |R_{I}\rangle - sin \gamma (cos \beta |G\rangle
+ sin \beta |R_{II}\rangle) \nonumber \\
\Psi_1 & = & sin \gamma |R_I \rangle + cos \gamma ( cos \beta |G\rangle
+ sin \beta |R_{II}\rangle )
\end{eqnarray}

If we ignore mass and phase space effects, and any differences between
the $n\bar{n}$ and $s\bar{s}$ wavefunctions, then proceeding as in the
previous section we obtain
\begin{eqnarray}
\Gamma((\Psi_1 + \Psi_2) \to \gamma\gamma) & = & \Gamma(q\bar{q} \to
\gamma\gamma) \times (\frac{1}{6} cos^2(\theta -\tau) + \frac{1}{6}
sin^2 \beta sin^2 (\theta - \tau)) \nonumber \\ 
\Gamma(\Psi_3 \to \gamma \gamma) & = & \Gamma(q\bar{q} \to
\gamma\gamma) \times \frac{1}{6} sin^2 (\theta - \tau) cos^2\beta
\end{eqnarray}
Defining $\Gamma_{\gamma \gamma} \equiv \Sigma_{i=1}^3 \Gamma(\Psi_i
\to \gamma\gamma)$ and later $\Gamma_{gg}$ analogously, the
generalisation of eq.(\ref{sum1}) becomes
\begin{equation}
\Gamma_{\gamma \gamma} = \frac{1}{6} \Gamma (q\bar{q}
\to \gamma\gamma).
\end{equation}
The generalisation of the relation for the gluon couplings,
eq(\ref{sum2}), becomes 
\begin{equation}
\Gamma_{gg} = 3\Gamma(q\bar{q} \to gg) + \Gamma(G \to gg).
\end{equation}
Consequently
\begin{equation}
\Gamma_{\gamma \gamma} = \frac{\alpha^2}{4 \alpha_s^2}
(1+c\frac{\alpha_s}{\pi})^{-1} (\Gamma_{gg}
- \Gamma(G \to gg) ).
\end{equation}
Thus, specialising to $0^{++}$ mesons, the experimentally measurable
quantities $\Gamma_{\gamma \gamma}$ and $\Gamma_{gg}$ obey the relation
\begin{equation}
\label{ineqgamgamfrgg}
\Gamma_{\gamma\gamma}[{\rm keV}] \leq \frac{(0.5/\alpha_s)^2}{20(1+ 8.6 
\frac{\alpha_s}{\pi})} \Gamma_{gg}[~{\rm MeV}].
\end{equation}
A major uncertainty comes from the large higher order QCD correction for
the $0^{++}$ sector which reduces the right hand side by a factor
of approximately 2.2
To be
conservative we therefore work to leading order.  If $f_0(1370)$ is
one of the trinity of glue associated states, then we infer from
$\Gamma(f_0 \to \gamma \gamma) = 5.4 \pm 2.3$ keV\cite{pdg94} that
\begin{equation}
\Gamma_{gg} \geq 108 \pm 46~ {\rm MeV}
\end{equation}
or from eq(\ref{0++}), neglecting mass dependence:
\begin{equation}
br(J/\psi \to \gamma + \Sigma_{i=1}^3 f_0^i) \geq (1.1 \pm 0.5)
10^{-3}.
\label{widthbound}
\end{equation}
Since $br(J/\psi \to \gamma f_J(1710)) \to K
\bar{K} = 0.97 \pm 0.12 ~10^{-3}$\cite{pdg94}, this bound is
satisfied by present data if the 1710 has $J=0$, even if $br(f_0(1710)
\rightarrow K \bar{K}) \sim 1$ and there is negligible production of
$f_0(1500)$ in radiative decay.  However the limit is only
barely respected so unless $\Gamma(f_0(1500) \to \gamma \gamma) +
\Gamma(f_0(1710) \to \gamma \gamma)$ is very small, either $br(J/\psi
\gamma f_0(1500))$ and or $br(J/\psi \gamma f_0(1370))$ must be
non-negligible, or $br(f_0(1710) \rightarrow K \bar{K}) < 1$.  If the
1710 state proves to have $J=2$, the bound (\ref{widthbound}) will be
very stringent indeed.  We now consider specific examples of mixing in
the $f_0(1370;1500;1710)$ system.  

\section{Three-State Mixings}
\label{3mixing}
\hspace*{2em}
An interesting possibility is that three $f_0$'s in the $1.4-1.7$ GeV 
region are admixtures of the three isosinglet states $gg$, $s\bar s$, and
$n\bar n$\cite{cafe95}.  Recently there have been two specific schemes
proposed which are based on lattice QCD and the emergent phenomenology
of scalar mesons.  In this section we present a simplified formalism
for treating a three component system of this type. 

At leading order in the glueball-$q \bar{q}$ mixing, ref\cite{cafe95}
obtained 
\begin{eqnarray}
\label{mixing}
N_G|G\rangle = |G_0\rangle + \xi ( \sqrt{2} |n\bar{n}\rangle + \omega |s\bar{s}
\rangle) \nonumber \\
N_s|\Psi_s\rangle = |s\bar{s}\rangle - \xi \omega |G_0\rangle \nonumber \\
N_n|\Psi_n\rangle = |n\bar{n}\rangle - \xi  \sqrt{2} |G_0\rangle
\end{eqnarray}
where the $N_i$ are appropriate normalisation factors, $\omega \equiv
\frac{E(G_0) - E(d\bar{d})}{E(G_0) - E(s\bar{s})}$ and the mixing parameter
$\xi \equiv \frac{\langle d\bar{d}|V|G_0\rangle}{E(G_0) - E(d\bar{d})}$. Our 
analysis suggests that the $gg \to q\bar{q}$ mixing amplitude manifested
in $\psi \to \gamma R(q\bar{q})$ is $O(\alpha_s)$, so that qualitatively
$\xi \sim O(\alpha_s) \sim 0.5$. Such a magnitude implies significant mixing
in eq.(\ref{mixing}) and is better generalised to a $3 \times 3$ mixing matrix.
Mixing based on lattice glueball masses lead to two classes of solution
of immediate interest:  

\noindent (i)$\omega \leq 0$, corresponding to $G_0$ in the midst
of the nonet\cite{cafe95} 

\noindent (ii)$\omega > 1$, corresponding to $G_0$ above the
$q\bar{q}$ members of the nonet\cite{wein96}. 

We shall denote the three mass eigenstates by $R_i$ with $R_1=f_0(1370)$,
$R_2=f_0(1500)$ and $R_3=f_0(1710)$, and the three isosinglet states 
$\phi_i$ with $\phi_1=n\bar n$, $\phi_2=s\bar s$ and $\phi_3=gg$ so
that $R_i=f_{ij}\phi_i$.  Recent data on the decay $f_0(1500) \to
K\bar{K}$\cite{landua} may be interpreted within the scheme of
ref\cite{cafe95} as being consistent with the parameter $\omega \sim
-2$. This enables simple analysis; if for illustration we adopt $\xi
=0.5 \sim \alpha_s$, the resulting mixing amplitudes are (scheme ``$A$''):
$$
\begin{array}{c c c c}
&f_{i1} & f_{i2} & f_{i3} \\
f_0(1370) & 0.86 &  0.13 & -0.50\\
f_0(1500) & 0.43 & - 0.61 & 0.61\\
f_0(1710) & 0.22 & 0.76 & 0.60\\
\end{array}
$$
By contrast, Weingarten\cite{wein96} has considered the case where the 
bare glueball lies above the $s\bar{s}$ member of the nonet. His
mixing matrix is (scheme ``$B$''):
$$
\begin{array}{c c c c}
&f_{i1} & f_{i2} & f_{i3} \\
f_0(1370) & 0.87 &  0.25 & -0.43\\
f_0(1500) & -0.36 &  0.91 & -0.22\\
f_0(1710) & 0.34 & 0.33 & 0.88\\
\end{array}
$$ 

The solutions for the lowest state are similar, as are the relative
phases and qualitative importance of the $G$ component in the high
mass state.  Both solutions exhibit destructive interference between
the $n\bar{n}$ and  $s\bar{s}$ flavours for the middle state.  

If we make the simplifying assumption that the photons couple to the
$n\bar{n}$ and  $s\bar{s}$ in direct proportion to the respective
$e_i^2$ (i.e. we ignore mass effects and any differences between the
$n\bar{n}$ and $s\bar{s}$ wavefunctions), then the corresponding two
photon widths can be written in terms of these mixing coefficients:
\begin{equation}\label{mixings}
\Gamma(R_i)=|f_{i1}\frac {5}{9\sqrt{2}}+f_{i2}\frac {1}{9}|^2 \Gamma,
\end{equation}
where $\Gamma$ is the $\gamma\gamma$ width for a $q\bar q$ system
with $e_q=1$.  One can use eq. (\ref{mixings}) to evaluate the
relative strength of the two photon widths for the three $f_0$ 
states with the input of the mixing coefficients.  These are (ignoring
mass dependent effects) 
\begin{equation} \label{acgamma}
f_0(1370) : f_0(1500) : f_0(1710) \sim 12:1:3
\end{equation}
in scheme $A$, to be compared with
\begin{equation} \label{weingamma}
f_0(1370) : f_0(1500) : f_0(1710) \sim 13:0.2:3
\end{equation}
in scheme $B$.  At present the only measured $\gamma \gamma$ width
in this list is that of the $f_0(1370) = 5.4 \pm 2.3$ keV\cite{pdg94}.
Using this to normalise the above, we anticipate $f_0(1500) \to \gamma
\gamma \sim 0.5$ keV (scheme $A$) or $\sim 0.1$ keV (scheme
$B$). Both schemes imply $\Gamma(f_0(1710) \to \gamma \gamma) =
1-2$ keV. 

This relative ordering of $\gamma \gamma$ widths is a common feature
of mixings for all initial configurations for which the bare glueball
does not lie nearly degenerate to the $n\bar{n}$ state.  As such, it
is a robust test of the general idea of $n\bar{n}$ and $s\bar{s}$
mixing with a lattice motivated glueball.  If, say, the $\gamma
\gamma$ width of the $f_0(1710)$ were to be smaller than the
$f_0(1500)$, or comparable to or greater than the $f_0(1370)$, then
the general hypothesis of significant three state mixing with a
lattice glueball would be disproven. The corollary is that qualitative
agreement may be used to begin isolating in detail the mixing pattern. 

Now we turn to $J/\psi$ radiative decay rates.  Since in either scheme
\begin{equation}
\Gamma_{\gamma \gamma} = 7.5 \pm 2.8~ {\rm keV},
\end{equation}
the discussion of the previous section implies,
\begin{equation}
br(J/\psi \to \gamma \Sigma f_0) \geq (1.5 \pm 0.6) \times 10^{-3}.
\end{equation}
However each scheme makes a more specific prediction.  By our
hypothesis that $q\bar{q}$ coupling to $gg$ is suppressed at 
$O(\alpha_s)$ relative to the corresponding glueball amplitude, we may
scale the $J/\psi \to \gamma f_0$ production amplitudes for the mixed
states as follows. For simplicity we shall assume that $A(gg \to
n\bar{n}) = \sqrt{2} A(gg \to s\bar{s}) = c \alpha_s A(gg \to G)$,
where $c$ is some constant whose magnitude and phase are in general
model dependent.  In this approximation, we have for scheme $A$ 
$$
\begin{array}{c c c}
A(f_0(1370) \to gg) & = & (-0.5 + c\alpha_s 1.3)A_0\\
A(f_0(1500) \to gg) & = &  0.6 A_0                \\
A(f_0(1710) \to gg) & = &  (0.6 + c\alpha_s 1.1) A_0\\
\end{array}
$$ 
In general we see that for mixing scheme $A$:

\noindent (i)The absence of a dominant signal in $J/\psi \to \gamma
f_0(1370)$ suggests that $c$ is not negative and that the
$G$-$q\bar{q}$ interference there is destructive. 

\noindent(ii)The $q\bar{q}$ admixture in the $f_0(1500)$ is nearly
pure flavour octet and hence decouples from $gg$. This leaves the
strength of  $br(J/\psi \to \gamma f_0(1500))$ at about $40\%$ of the
pure glueball strength, which is consistent with the mean of the two
analyses in section \ref{ss:f1500}. 

\noindent(iii)The destructive interference in the $f_0(1370)$ case
implies a constructive effect for the $f_0(1710)$ and hence this picture
predicts that $ br(J/\psi \to \gamma f_0(1710)) >br(J/\psi \to \gamma
f_0(1500)) > br(J/\psi \to \gamma f_0(1370))$.  If 
as a particular example for comparison between the two schemes
we take $c \alpha_s = 0.5/1.3$ to decouple $f_0(1370)$ entirely in
radiative $J/\psi$ decay, we find $ br(J/\psi \to \gamma
f_0(1710)):~br(J/\psi \to \gamma f_0(1500):~ br(J/\psi \to \gamma
f_0(1370)) =  1.1:~0.4:~0 $. 

Mixing scheme $B$, corresponding to an ideal glueball lying above the
nonet, leads to the following amplitudes: 
$$
\begin{array}{c c c}
A(f_0(1370) \to gg) & = & (-0.4 + c\alpha_s 1.5) A_0\\
A(f_0(1500) \to gg) & = & (-0.2 + c\alpha_s 0.4)A_0\\
A(f_0(1710) \to gg) & = & ( 0.9 + c\alpha_s 0.8)A_0\\
\end{array}
$$
Here both $f_0(1370)$ and $f_0(1500)$ production are suppressed
due to the destructive interference of the glueball and $q\bar{q}$
components; the $f_0(1710)$ being enhanced as in the previous example.
For the example $c \alpha_s = 0.4/1.5$, (chosen to decouple the
$f_0(1370)$ and enable comparison with scheme $A$ as above) we find $
br(J/\psi \to \gamma f_0(1710)):~br(J/\psi \to \gamma 
f_0(1500):~ br(J/\psi \to \gamma f_0(1370)) = 1.2:~0.01:~0$. 

Thus, in conclusion, both these mixing schemes imply a similar hierachy
of strengths in $\gamma \gamma$ production which may be used as a 
test of the general idea of three state mixing between glueball and
a nearby nonet. Prominent production of $J/\psi \to \gamma f_0(1710)$
is also a common feature.  When the experimental situation clarifies
on the $J/\psi \to \gamma f_0$ branching fractions, we can use the
relative strengths to distinguish between the case where the glueball
lies within a nonet, ref\cite{cafe95}, or above the $s\bar{s}$ member,
ref\cite{wein96}. 

\section {Summary}
\hspace*{2em}
We have clarified the relationship between $b_{rad}({Q \bar{Q}}_V \rightarrow \gamma +R)$ and $br(R \rightarrow gg)$
proposed by Cakir and Farrar\cite{cak}.  In particular, we have
examined its dependence on the $<g g |R>$ form factors and 
discussed theoretical and experimental constraints on these form
factors.  We conclude that the relation can be used, possibly with
generalized $H_J(x)$ functions, for light-$q \bar{q}~$ mesons and glueballs
as well as heavy $q \bar{q}~$ mesons.  Using this relation, we find
\begin{itemize}
\item  The $f_0(1500)$ is at least half-glueball if the Bugg et al
analysis\cite{bugg} of the $4 \pi$ channel is confirmed,  but is less
so according to the BES results.  Analysis of MarkIII data on $J/\psi \to
\gamma \pi \pi$ is urgently needed.  At this moment the experimental
determinations of $\Gamma(J/\psi \to \gamma f_0(1500))$ are inconsistent.
\item  The $f_J(1710)$ is also at least half-glueball, if $J=0$; if 
 $J=2$ it is a $q \bar{q}~$ meson.  Experimental determinations of the
$f_{0,2}$ spectra in the $1.6-1.8$ GeV region are presently inconsistent.
\item  The $\xi(2330)$ is unlikely to have $J=0$, if present
experimental data are correct.  If it has $J=2$ it strongly resembles
a glueball.
\item The $\eta(1440)$ is separated into two states. The lower
mass state, $\eta_L(1410)$, has strong affinity for glue; the higher mass
$\eta_H(1480)$ is consistent with being 
the $s\bar{s}$ member of a nonet, perhaps
mixed with glue.
\end{itemize}
It is of urgent importance to (a) arrive at an experimental consensus
on the $f_0$ and $f_2$ masses and widths in the 1600-1800 region and
(b) resolve the discrepancies in the present determinations of
$br(\psi \to \gamma f_0(1500)) $.  Measurement of production branching
fractions of the $f_0$ and $f_2$ mesons in $\Upsilon$ radiative decay
should be quite easy and yield useful additional information.
We also outlined a procedure to use data on $\psi \to \gamma R$ and
$\gamma \gamma \to R$ together, to help unravel the $q \bar{q}$ and
$gg$ composition of mesons.  To accomplish this, measurement of
$\Gamma(f_0(1370; 1500; 1710) \to \gamma \gamma)$ is an essential
ingredient. 

An emerging mystery is the $\eta(1440)$ region.
Its properties seem to differ in $J/\psi$ radiative decay and
$p\bar{p}$ annihilation, and it has not been seen in central
production.  Possibly these differences are due to the different
interplay of gluon and $q\bar{q}$ annihilation in the various
production processes. This merits further investigation, both
experimental and theoretical.  The strong production of the
$\eta(1410)$ in radiative $J/\psi$ decay indicate that it could be a
glueball.  However its low mass is difficult to reconcile with lattice
gauge predictions.  Its properties and mass are consistent with those
expected for a bound state of light gluinos.  Given that the
$\eta(1410)$ may be evidence of a new degree of freedom in QCD, or
evidence of dynamics beyond quenched lattice gauge theory in the
$0^{-+}$ sector, more detailed experimental investigation of the
pseudoscalar sector is a high priority.

\section{Acknowledgements}
\hspace*{2em}
We are indebted to D.V.Bugg, W. Dunwoodie and A.Kirk for discussions.
This work supported by U.S. National Science Foundation grant nos. 
NSF-PHY-94-23002 (GRF) and NSY-PHY-90-23586 (ZPL). FEC is partially
supported by the European Community Human Mobility Program Eurodafne,
Contract CHRX-CT92-0026

\section*{Appendix: Analytical Expressions for $H_J(x)$}
The analytical expressions for the loop integral $H_J(x)$ are given in
Ref. \cite{korner}. In the normalisation of the present paper they are:
\begin{eqnarray}\label{hj0-+}
H_{0^{-+}}(x)=\frac 4x \left [L(1-2x)-L(1)
     -\frac {1-x}{2-x}(2L(1-x)-\frac {\pi^2}3+
     \frac 12\ln^2(1-x))\right . \nonumber \\ \left . 
-\frac x{1-2x}\ln(2x) \right ]
+i4\pi\frac {1-x}{(2-x)x}\ln(1-x)
\end{eqnarray}
for $J=0^{-+}$, and 
\begin{eqnarray}\label{hj0++}
H_{0^{++}}=\sqrt{\frac 23} \left [
\frac {2-3x}{x^2}+\left (10\frac {1-x}{x^3}
+4\frac {1-2x}{x^2}\ln(2)\right )\ln(1-x) \right .
\nonumber \\
+\left (\frac 8{x^2}+2\frac{1 -x}{x(1 -2x)}\right )\ln(2x) .
-3\frac {1 -x}{x(2-x)}\ln^2(1 -x) \nonumber \\
+\frac {8-6 x+x^2 -6x^3}{x^3(2 -x)}\pi^2/6  
-\frac {4 -5x+2x^2}{x^3}L(1 -2x) \nonumber \\ \left .
 - 4\frac {2-2x-x^2}{x^2(2-x)}L(1 -x)+ i\pi 6
 \frac {1-x}{x(2 -x)}\ln(1 -x) \right ]
\end{eqnarray}
for $0^{++}$, 
where $L(x)$ is a Spence function, defined as
\begin{equation}\label{spence}
L(x)=-\int_0^x \frac {dx}{x} \ln(1-x).
\end{equation}
There are  three helicity amplitudes for the tensor state, and 
they are related to the total $H_{2^{++}}$ by
\begin{equation}\label{hj2++}
|H_{2^{++}}(x)|^2=|H_{2^{++}}^0(x)|^2+|H_{2^{++}}^1(x)|^2
+|H^2_{2^{++}}(x)|^2.
\end{equation}
The helicity amplitudes $H_{2^{++}}^l$  in Eq. \ref{hj2++} are
\begin{eqnarray}
H_{2^{++}}^0=\frac {2\sqrt{3}}{x^3} 
\left [ x(6-5x)+ \frac 23 \frac {6 -19x+18x^2}x (1 -x)\ln(1 -x) \right . 
\nonumber \\
-\frac {10-12x+5x^2}{3(2 -x)} g_1 
+\frac 23 \frac {6-38x+71x^2 -37x^3}{1-2x}\ln(2x)
\nonumber \\
- 8\frac {(1-x)^2}
{x^2(2 -x)}g_2+\frac 43 \frac {6 -6  x-x^2}x\left (\ln(2)-\frac 12 i \pi
\right ) \nonumber \\ \left . -\frac 43  (12 -26  x+13  x^2 ) g_3 \right ],
\end{eqnarray}

\begin{eqnarray}
H_{2^{++}}^1= \frac {2\sqrt {1-x}}{x^3} \left [ 
-\frac 13 (38 -9  x) x -\frac 2x (4-13 x+16 x^2
-4x^3) \ln(1 -x) \right .
\nonumber \\ 
-2\frac {x (1 -x)}{2 -x} g_1  
-\frac 4{1-2x}(2 -11  x+16x^2 -4x^3)\ln(2x)
\nonumber \\
+8  \frac {(1 -x) (2 -2  x+x^2)}{x^2(2-x)} g_2
-\frac {16}3 \frac {3 -3  x+x^2}{x} \left (\ln(2)-\frac 12 i\pi\right )
\nonumber \\ \left .
 +4(8 - 12x+3x^2)g_3\right ]
\end{eqnarray}
and 
\begin{eqnarray}
H_{2^{++}}^2= \frac {\sqrt{2}(1-x)}{x^3} 
\left [ \frac {16}3 x + \frac 4x  (1-6x+6x^2)\ln(1-x)+
2  \frac {5 -6x+2x^2}{2-x} g_1 \right . \nonumber
  \\
+4(1-6x) \ln(2x)-4  
\frac {2 -4x+6x^2 -4x^3 +x^4}{x^2(2-x) } g_2
\nonumber \\ \left .
+\frac 43\frac {6 -6x+11x^2}{x}\left (\ln(2)-\frac 12 i\pi\right ) 
-16(1-x)g_3 \right ]  
\end{eqnarray}
where 
\begin{equation}
g_1=L(1)-L(1-2x),
\end{equation}

\begin{equation}
g_2=L(1-2x)-2 L(1 -x)+L(1)-\frac 12 \ln^2(1 -x) + i\pi \ln(1-x)
\end{equation}
and
\begin{equation}
g_3=L(1-x)-L(1-2x) -\ln(2) \ln(1-x)
\end{equation}


\begin{figure}
\epsfxsize=\hsize
\epsffile{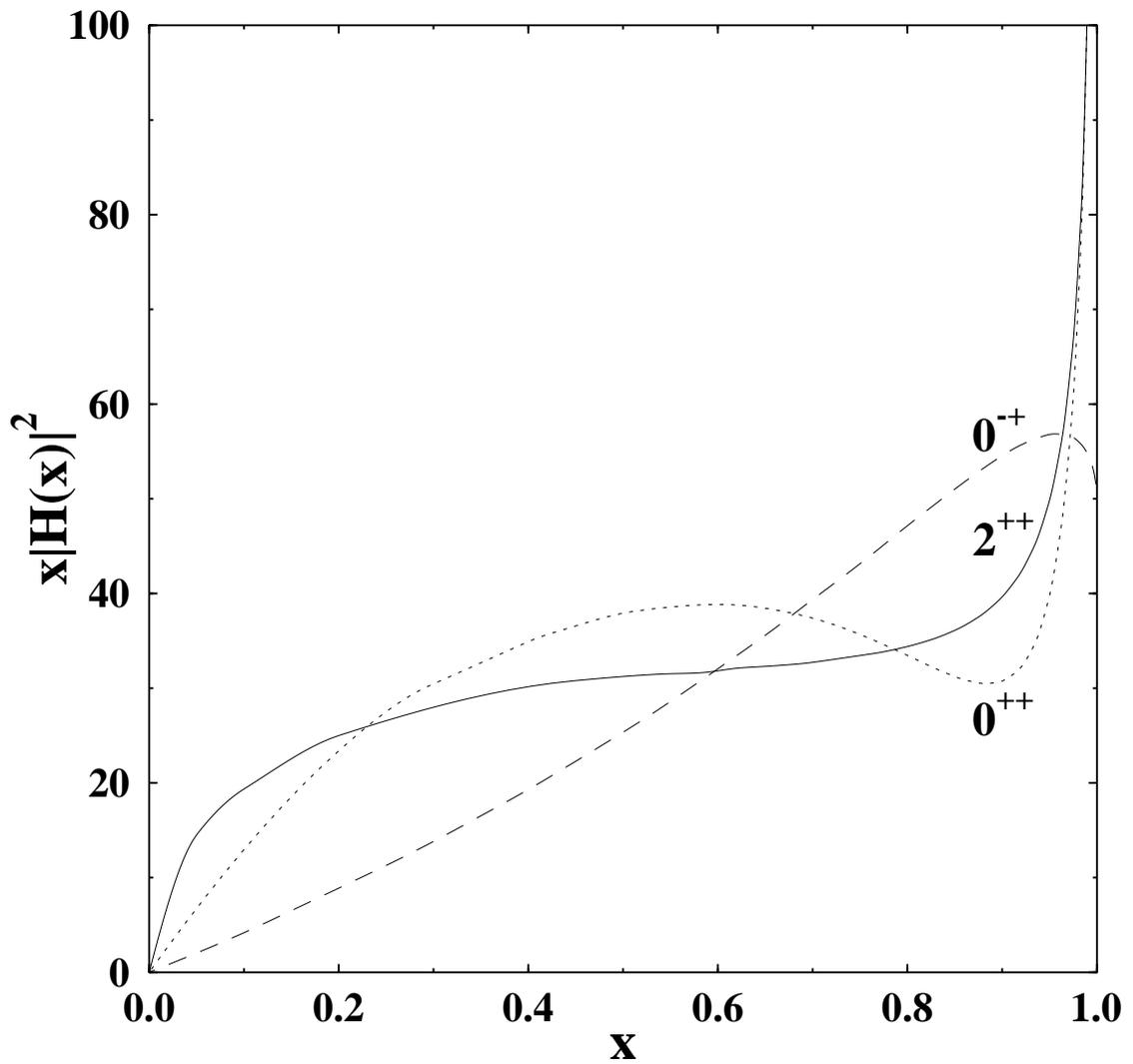}
\caption{Magnitude of the loop integral, $x|H|^2$ versus $x$ for $0^{++}$ 
(dotted), $0^{-+}$ (dashed) and $2^{++}$ (solid); $x=1-(\frac{m_R}{m_V})^2$.}  
\label{HJ}
\end{figure}

\end{document}